\documentclass{cernphprepmod}   
\usepackage{relsize}
\usepackage{hyperref} 
\usepackage{lineno}
\begin{document}
\begin{titlepage}
\PHdate{} 
\vspace{1.2cm}
\title{\LARGE Space charge in liquid argon time-projection chambers:\\  
a review of analytical and numerical models, \\
and mitigation methods}
\begin{Authlist}
Sandro Palestini and Filippo Resnati
\Instfoot{a1}{CERN, 1211 Geneva 23, Switzerland}

\end{Authlist}
%
%
\ShortAuthor{}
\ShortTitle{}
\begin{abstract}
The subject of space charge in ionisation detectors is reviewed, 
with particular attention to the case of liquid argon time projection chambers. 
Analytical and numerical description of the effects on the reconstructed coordinates
along the drift and the transverse directions are presented. 
The cases of  limited electron lifetime, 
of dual-phase detectors with ion feedback, 
and of detectors 
with small and comparable ratio between drift length and width are considered.
Two design solutions that mitigates the effects are discussed. 
\end{abstract}
\Submitted{(Submitted to Journal of Instrumentation)}
\end{titlepage}
%

%
\section{Introduction}
Ionisation detectors, based on gaseous or liquid media, are designed to collect the signal induced by the motion of electrons between polarisation electrodes. In comparison, the signal induced by ions is very low, because their drift velocity 
is lower by several orders of magnitude. 
However the low velocity also implies that the ions remain longer in the drift volume: in case of long drift distance or large irradiation, 
and in particular for liquid detectors, the charge density due to ions may be large enough to affect the electric field 
that drives the motion of the electrons.\footnote 
{On the other side, in gaseous detectors space-charge effects are usually due to ions from multiplication in gas rather than primary ionisation. 
The subject was discussed first in the context of gas diodes~\cite{SpaceCharge-hystorical-1,SpaceCharge-hystorical-2}, before becoming a subject of relevance in detectors for particle physics~\cite{SpaceCharge-gasTPC-1,SpaceCharge-gasTPC-2}.
 }

Effects of space charge have been studied in calorimeters using liquid krypton~\cite{Palestini:1998an} 
and liquid argon~\cite{LArCalorimeters, ion-mobility-critical}. More recently, the relevance and the scope of this subject have widened with the development of large liquid argon imaging devices~\cite{Mooney:2015kke, Abratenko:2020bbx,Antonello:2020qht,ProtoDUNE-SP,ProtoDUNE-DP}, where the drift volume may be long enough to become sensitive to the space charge due just to the exposure to cosmic rays at the surface of the Earth.

The purpose of this paper is to review the basic formalism describing space-charge effects in ionisation detectors. 
First, the simplest one-dimensional treatment will be reviewed, followed by a discussions of the effects of charge-yield dependence on the electric field, and of electron attachment, with the related presence of negative ions. 
Multi-dimensional cases are discussed in terms of boundary conditions on the side walls, and of aspect ratio (depth vs.\ width) of the drift volume.  
The case of dual-phase detector with injection of positive ions at the anode end of the drift volume is discussed. 
Two mitigation methods are considered, dealing respectively with the effects on the drift coordinate and on the transverse coordinates. Throughout the work, scaling laws are discussed, aiming at relations that could be used as guidelines for detectors operated in different conditions and with different geometry. Comparison between analytical approximations and numerical solutions are presented.  

\section{Review of basic assumptions and simplest case}   \label{sec:1D-model}
This section will first summarise the results of \cite{Palestini:1998an}, dealing with the simplest, one-dimensional equation for space charge and electric field. 
Next, it will consider aspect related to an ionisation detector used as a time projection chamber (TPC).

\subsection {One-dimensional analysis} 
An ionising particle creates pairs of electrons and positive ions, but because the respective values of drift velocity differ by typically by 5 or 6 orders of magnitude, under steady conditions the density of the two charge carriers differ by the same amount. 
Therefore, in the fundamental approximation the density of charge is assumed to be due 
to positive ions alone, $\rho^+$, which varies because of the injection of ions from ionising particles, and satisfies the continuity equation
\begin{equation}
\frac{\partial \rho}{\partial t}^+ +\overline \nabla\cdot(\rho^+ \overline v^+) = K \; .
\end{equation}
The stationary solution ($\partial \rho^+\!/\partial t=0$) is of interest, and it is determined under the assumption of constant and uniform charge density injection rate $K$. The value of $K$ depends on the flux of ionising particles crossing the detector and on the value of the electric field, which affects the initial recombination of electrons and ions. 
The space charge causes a non uniformity in the electric field, and therefore a space dependence of $K$,
which is ignored here, and considered below in section~\ref{sec:recomb}. 
The assumption of uniformity and stability of $K$ is valid because the charge injection, due to e.g., cosmic rays, 
is effectively averaged over the time needed for the ions to drift from anode to cathode.
The detector is taken as a parallel-plates ionisation chamber with gap $L$, operated with voltage $V_\circ$ and 
average electric field $E_\circ=V_\circ/L$ directed along $+x$, with $x=0$ ($x=L$) at the anode (cathode).
The problem is reduced to one-dimension under the assumption that far from the side wall the boundary effects are 
negligible, so that the dependence on $y$ and $z$ may be ignored. 

Under these assumptions, and using the mobility $\mu^+$, 
the continuity equation 
\begin{equation} 
\frac{d (\rho^+ v_x\!^+)}{dx} = K \;   \label{eq:continuity}
\end{equation}
is solved as 
\begin{equation}
\rho^+\!(x) = \frac{K\, x}{\mu ^+ E_x(x) }\; , \label{eq:density+}
\end{equation}
where $v_x^+ = \mu ^+ E_x$ is used and the boundary condition $\rho^+ (0)=0$ 
is applied, since the positive ions drift away from the anode and no accumulation of space charge is 
possible at $x=0$. For $x > 0$, space charge is present reflecting 
the rate of charge density injection, accumulated over a time effectively equal to $x/(\mu ^+ E_x)$. 

The electric field satisfies the Gauss's law, which under these assumptions is written as 
\begin{equation}
\frac{dE_x}{dx} =  \frac{\rho^+}{\epsilon}  = \frac{K \, x}{\epsilon \, \mu ^+ E_x } \label{eq:gauss} 
\end{equation}
and is solved directly as 
\begin{equation}
E_x(x) =  E_\circ \sqrt{(E_\text a/E_\circ)^2 + \alpha^2 (x/L)^2 } \;,  \label{eq:Ex}
\end{equation}
where 
$E_\text a$ is the value of the 
electric field at the anode, which is determined by the boundary 
$\int\! E_xdx = V_\circ$ 
integrated from anode to cathode, 
while the dimensionless parameter $\alpha$ is defined as~\cite{Palestini:1998an} 
\begin{equation}
\alpha =  \frac{L}{E_\circ} \sqrt{\frac{K}{\epsilon \mu^+}} \;.  \label{eq:alpha}
\end{equation}
This parameter can be interpreted as $\alpha^2$ being equal to the charge density injection rate $K$,
multiplied by the gap length $L$ and by the ions drift-time across the gap  $L/(\mu^+ E_\circ)$, 
and divided by the surface charge density $\sigma_\circ = \epsilon E_\circ$ at the electrodes,
with the last two quantities computed for vanishing $K$. The ratio $\sigma_\circ/L\equiv \rho_\circ$ is the natural unit for evaluating 
$\rho^+(x)$, which can be written as
\begin{equation}
\rho^+\!(x) = \alpha^2 \,  \rho_\circ \,\frac{E_\circ}{E_x(x)} \;\frac{x}{L} \; .
\end{equation}
\begin{figure}[!bt]
\begin{center}
\includegraphics[width=0.7\textwidth]{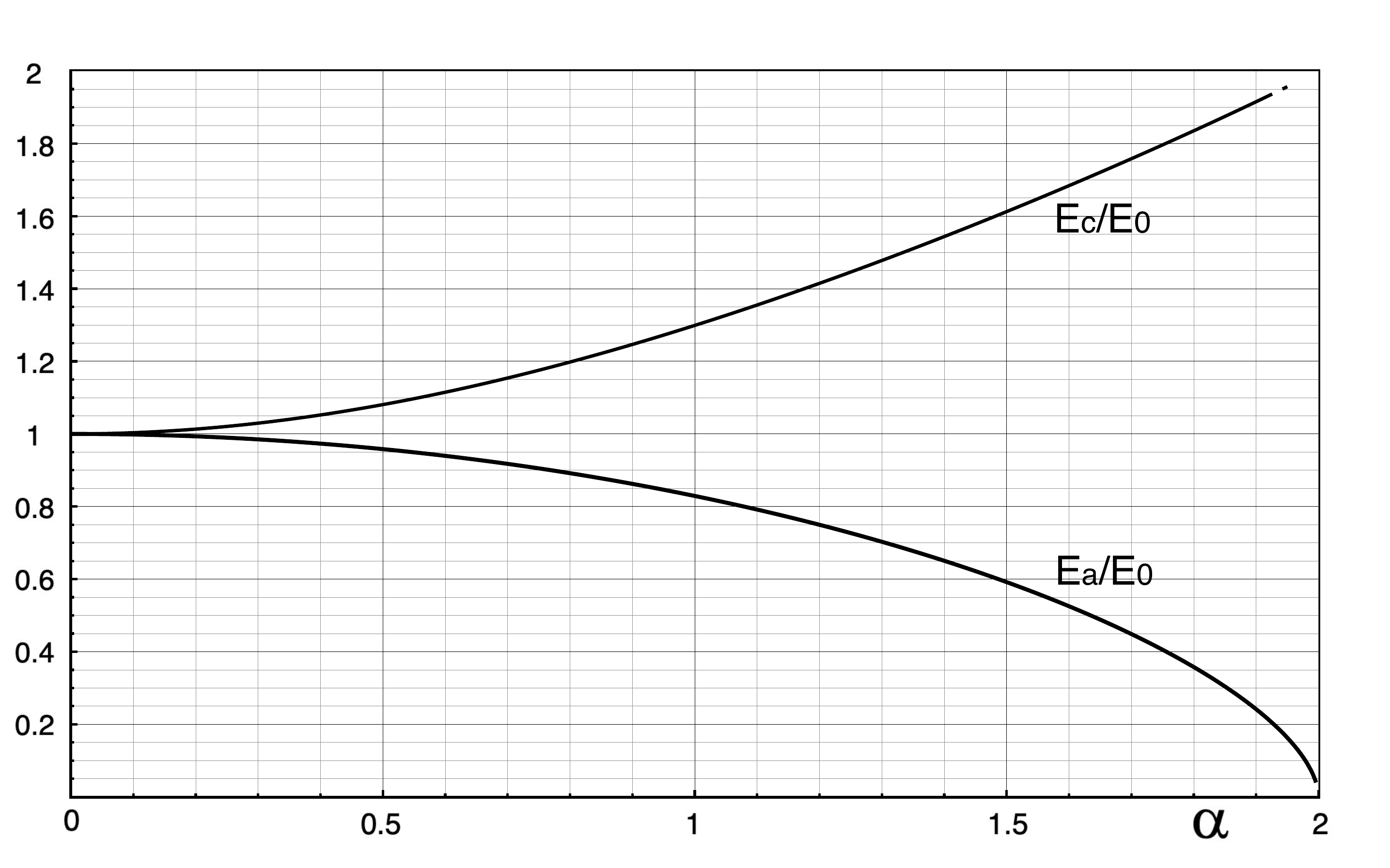}  
\caption {Normalized electric field at the anode $E_\text a / E_\circ$ and cathode $E_\text c / E_\circ$ 
as a function of the dimensionless parameter $\alpha$, from reference~\cite{Palestini:1998an}.}
\label{fig:Ea-Ec}
\end{center}
\end{figure}

From eq.~\ref{eq:Ex}, and as shown in figure~\ref{fig:Ea-Ec}, the field at the anode $E_\text a$ is always lower than $E_\circ$, while the opposite holds for the field at the cathode $E_\text c$.
Correspondingly, the electric potential, defined as $V(x) = -\int E_x \, dx$ and integrated from the anode towards the cathode, 
deviates from the value $-E_\circ\, x$, according to a smooth convex profile. 
The analytical expression for $V(x)$ is
given below in eq.~\ref{eq:V(x)-analytical}, appendix~\ref{app:analytical expressions}.

As discussed in \cite{Palestini:1998an},
 a critical situation occurs for $\alpha \ge 2$, 
 when the electric field vanishes at the anode, enhancing recombination between electrons and positive ions.\footnote
{In the critical condition $\alpha=2$ the electric field varies linearly as $E_x=2\, E_\circ (x/L)$ and the density 
of positive ions is uniform: $\rho_c=K\,L/(2\mu^+E_\circ)=2\, \rho_\circ$. 
For $\alpha >2$ the active region is reduced to a gap of length $L' = 2 L/\alpha$, detached from the anode by $L-L'$, 
with $E_x=2\, E_\circ (x'/L')$ for $x'=x-(L-L')>0$, while $E_x$ is highly suppressed for  $x \leq L-L'$.
The occurrence and the modality of a critical condition, predicted in~\cite{Palestini:1998an}, has been observed in calorimetric cells~\cite{ion-mobility-critical}.}

\begin{figure}[bt]
\begin{center}
\includegraphics[width=0.7\textwidth]{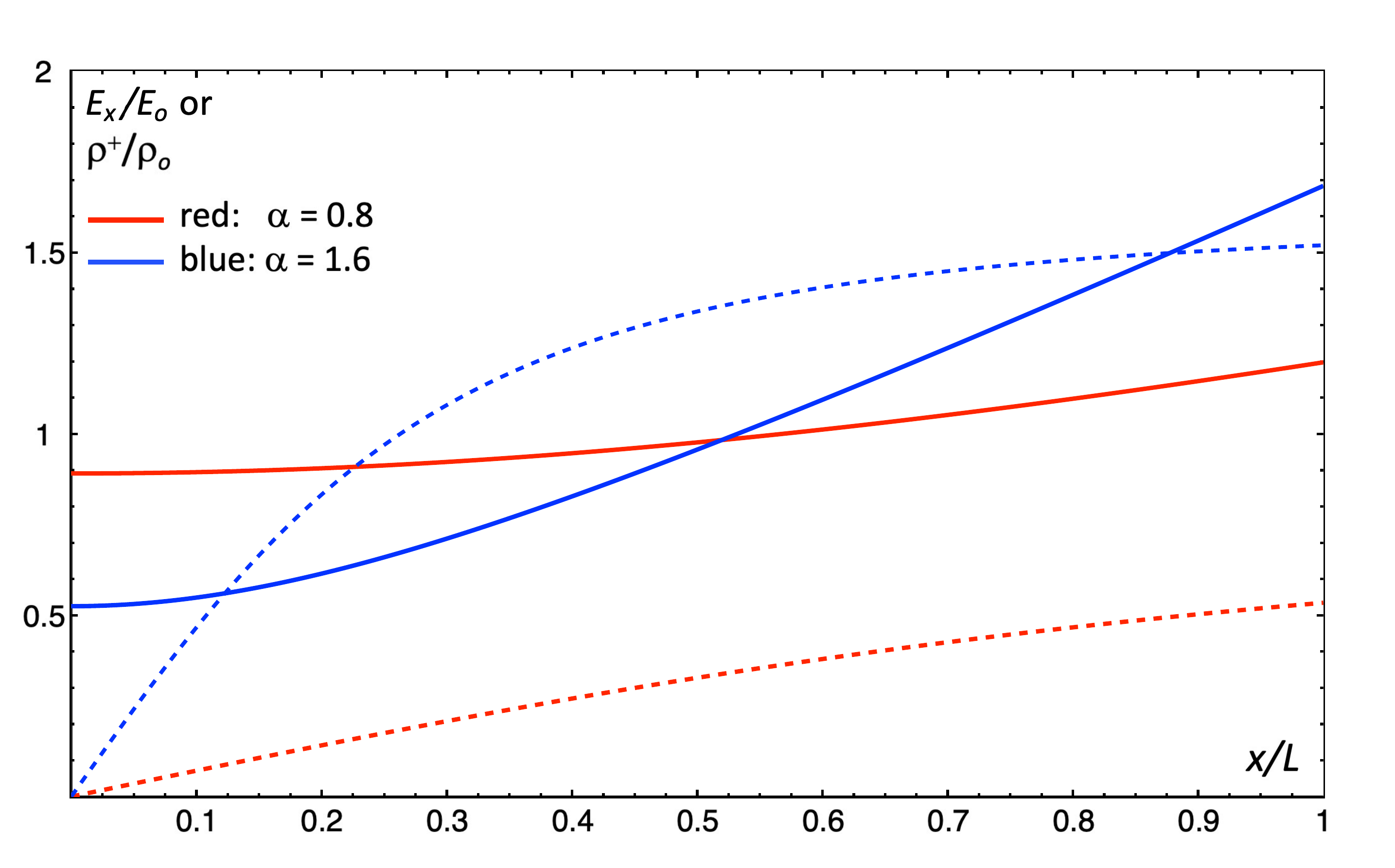}
\caption {Electric field (continuous lines) and charge density (dashed line) behaviour for $\alpha = 0.8$ (red) and $\alpha = 1.6$ (blue).
The horizontal axis is the drift coordinate divided by the gap length ($x/L$), with $x=0  \,(1)$ at the anode (cathode).  The electric field is in units of $E_\circ=V_\circ/L$, and the charge density in 
units of $\rho_\circ =\epsilon E_\circ/L$.
}
\label{fig:example-E-rho}
\end{center}
\end{figure}

As an example, in a liquid argon detector operated at the surface of the Earth, the charge density injection rate due cosmic rays 
is approximately given by $K=2\times10^{-10}$~C m$^{-3}$s$^{-1}$. 
For $L=4$~m, $E_\circ=500$~V/cm, dielectric constant of 1.504, 
and with $\mu^+=1.6\times 10^{-7}$~m$^2$V$^{-1}$s$^{-1}$ for the ion mobility, 
the dimensionless parameter takes the value $\alpha=0.78$. 
Figure~\ref{fig:example-E-rho} shows the behaviour of $E_x/E_\circ$ and $\rho / \rho_\circ$ vs. $x/L$ under similar 
assumptions. For comparison, the curves corresponding to $\alpha=1.6$ are also shown.

It should be kept in mind that there is uncertainty in the value of the mobility of Ar$^+$ ions, with reported values in the 
range of 0.8 to 2.0$\times 10^{-7}$~m$^2$s$^{-1}$V$^{-1}$ 
\cite{ion-mobility-critical,Antonello:2020qht,ion-mobility-1,ion-mobility-2,ion-mobility-3}, 
for values of temperature and pressure usual for particle detectors. Such uncertainties propagates directly to the value of the parameter $\alpha^2$, and to the application of the results 
presented in this paper. 

Appendix~\ref{app:analytical expressions} provides a set of relations valid in first-order expansion in the parameter $\alpha^2$, together with numerical approximations at higher order.
%
%

\subsection{Time projection chambers}
In a time-projection chamber, like liquid argon devices designed for neutrino detection~\cite{LArTPC-Rubbia}, the effect of the space charge on the collection time of a charge deposited at a distance $x$ from the anode 
is described by a time offset $\delta t$, given by 
\begin{equation}
\delta t (x)=  \int_0^x \left( \frac{1}{v\,^e\! (x')} - \frac{1}{v\,^e\!_\circ}\right) dx' \;,
\end{equation}
where $v\,^e\! (x)$, $v\,^e\!_\circ$ are the values of the drift velocity of electrons at the local electric field $E_x(x)$ and at the nominal value $E_\circ$, respectively.
Since the electron drift velocity depends monotonically on $E_x$, the integral receives contribution of different sign from regions of small and large $x$.
Using $t_\circ = L/v\,^e\!_\circ$ and taking $\delta v\,^e / v\,^e\!_\circ = \gamma (\delta E_x /E_\circ)$ to describe the 
non-linearity of the drift velocity of electrons, in the approximation of small values of $\alpha$ it is straightforward to find
\begin{equation}  \label{eq:delta-t}
\delta t(x) \simeq  \alpha^2  \frac{\gamma\;t_\circ}{6} \frac{x}{L} \left( 1-\frac{x^2}{L^2}\right)  \;,
\end{equation}
which naturally vanishes at $x=0$ and also at $x=L$, although the latter property holds only to first order in $\alpha^2$. 
The maximum effect on the drift time occurs for  $x_\mathrm {max} \simeq L/\sqrt{3}$, and is equal to  
\begin{equation}
\delta t_\mathrm {max} \simeq   \alpha^2  \frac{\gamma \; t_\circ}{9 \sqrt{3}} \; .
\end{equation}
The effect on the reconstructed coordinate, which can be referred to as \emph {longitudinal distortion}, is 
$\delta x  = v\,^e\!_\circ \, \delta t $, which in the same approximation has the largest value
\begin{equation}
\delta x _\mathrm {max}\simeq   \alpha^2  \frac{\gamma \, L}{9 \sqrt{3}}   =   \frac{\gamma}{9 \sqrt{3}}        \frac{L^3}{E_\circ\!^2} \frac{K}{\epsilon \mu^+} \; .
\label{eq:longitudinal-dist}
\end{equation}
This equation shows that the longitudinal distortion is proportional to the charge density injection rate, $L^3$, and $E_\circ{}^{-2}$. 
In the example considered above (liquid argon TPC with $\alpha=0.78$, $L=4$~m), and using $\gamma \simeq 0.5$~\cite{electron-velocity},
the maximum longitudinal distortion is $\delta x _\mathrm {max} \simeq 7.7$~cm.

The variation in the drift velocity due to space charge has a minor effect on the diffusion of the drifting electrodes.  The effect is maximum for $x \simeq x _\mathrm {max}$. 
In the same example, taking into account the partial compensation due to the dependence of the diffusion coefficient on the electric field \cite{diffusion}, the maximum increase in the longitudinal diffusion has the negligible value of about 0.1~mm.  


\subsection{Electric field variation and ionisation yield}  \label{sec:recomb}
\begin{figure}[tb]
\begin{center}
\includegraphics[width=0.7\textwidth]{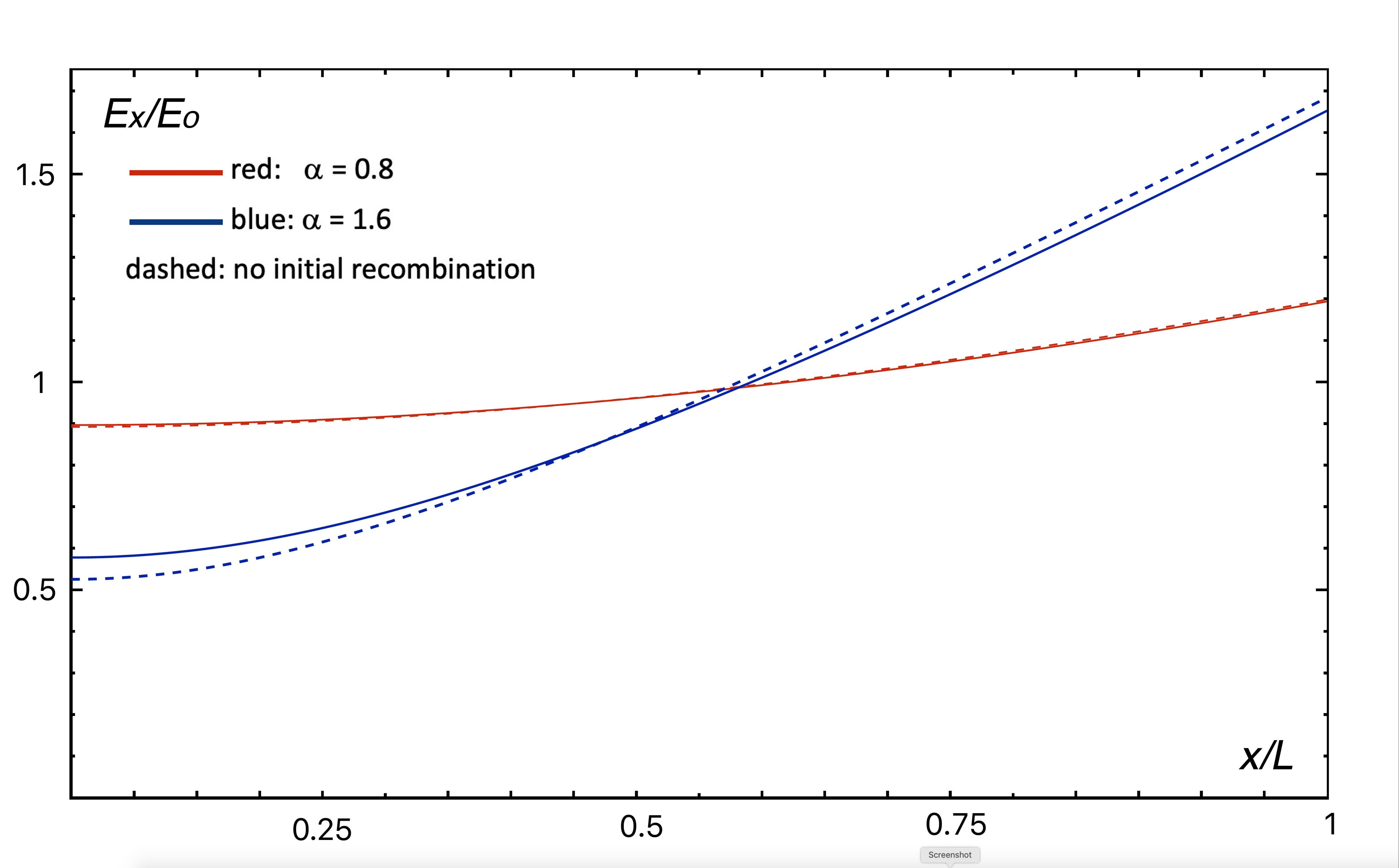}
\caption {Normalised electric field $E_x(x)/E_\circ$ vs.\ $x/L$ for $\alpha=$0.8, 1.6, taking into account recombination (continuous lines) and
ignoring recombination (dashed lines).
}
\label{fig:recombination}
\end{center}
\end{figure}

The analytical description contained in section~\ref{sec:1D-model} can be brought to a more realistic condition by introducing an $x$ dependence 
in the change density injection, related to initial charge recombination, 
which changes the amount of free 
electrons and 
ions as a function of the electric field strength. To this purpose $K$ is multiplied by $R(E)$ and, with $E=E_x$, eqs.~\ref{eq:continuity} 
and \ref{eq:Ex} are replaced by:
\begin{equation} 
\frac{d (\rho v_x\!)^+}{dx} = K\, R(E_x)\: ,   \qquad 
\frac{dE_x}{dx} =  \frac{\rho^+}{\epsilon} = \frac{K \int_0^x R(E_x(x'))dx'}{\epsilon \, \mu^+ E_x(x)\, }\; .
\end{equation}
Figure~\ref{fig:recombination} shows examples of numerical solutions for $E_x(x)/E_\circ.$
$R(E_x)$ is taken from \cite{recombination} and normalised to the value for $E_\circ=500$~V/cm: $R(E_x)=1.15/(1+72.9/E_x)$, with $E_x$ expressed in V/cm. 
Values $\alpha=0.8$ and 1.6 are considered, including recombination (continuous lines) and
excluding it (dashed lines). The difference is negligible for the smaller value of $\alpha$, and rather small for the larger.  The effects of recombination would be more visible for larger values, and the threshold for critical density is increased to $\alpha \simeq 2.5$. 

Besides the effect on $\delta x$, the $x$ dependence of the field strength needs to be taken into account when the specific energy loss $dE/dX$ along the trajectory of a charged track is extracted from the ionisation signal $dQ/dX$,
for the purpose of particle identification. The local dependence of the charge yield affects the ionisation signal as
$dE \propto dQ / R(E(x))$. Furthermore, the segment on the track length  $dX = (dx^2 + dy^2 + dz^2)^{0.5} $ requires a
\emph {scale correction} for the $dx$ component, proportional to  $d(\delta x)/dx$. Using eq.~\ref{eq:delta-t} and  $\delta x  = v\,^e\!_\circ \, \delta t $, at first order in $\alpha^2$ the correction varies between $+0.08\, \alpha^2\cos ^2\theta_x$ at the anode to $-0.17\, \alpha^2\cos ^2\theta_x$ at the cathode, where $\theta_x$ is the angle between the track segment and the drift direction.

In the following sections, analytical results are obtained ignoring the dependence of the initial recombination 
on the electric field, while the effect is included in numerical computations.
%
%

\subsection{Electron attachment  and negative ions} \label{sec:electron-lifetime}
Drifting electrons may be captured by electronegative impurities according to  $d\rho_\text{e}/dt= \rho_\text{e}/\tau_\text{e}$.
The electron lifetime $\tau_\text{e}$ is related to the attachment rate constant $k_\text{e}$ and to the density of electronegative impurities $n_\text{i}$ as $\tau_\text{e}=1/(k_\text{e}\,n_\text{i})$. 
The attachment rate constant depends on the type of impurity, and on the electron energy distribution, which 
is affected by the electric field, but this dependence has been observed to be rather small for field strength below 
1000~V/cm~\cite{lifetime-1,lifetime-2}.

\begin{figure}[t]
\begin{center}
\includegraphics[width=0.7\textwidth]{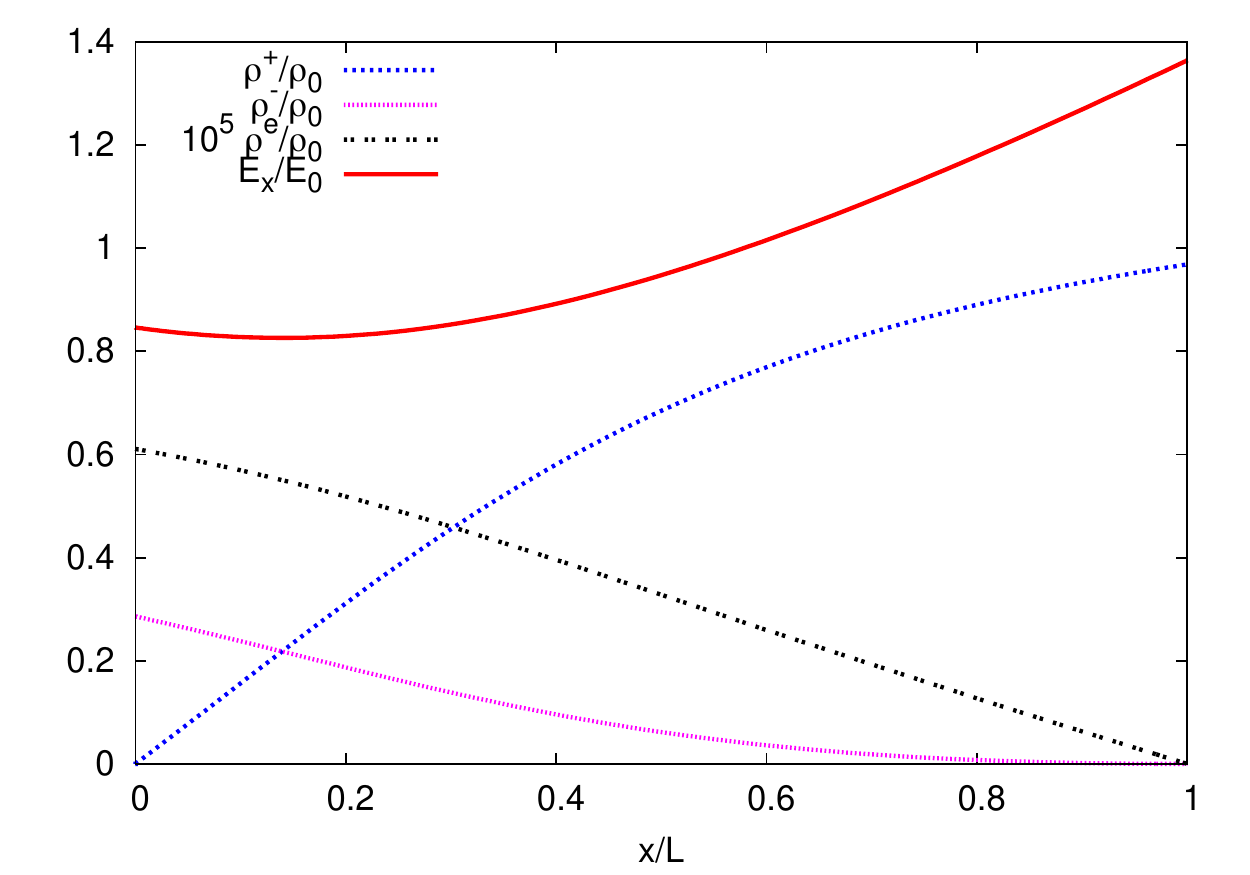} 
\caption {One-dimensional numerical computation of the effects of electron attachment. 
The parameters used are $L = 6$ m, $\alpha=1.15$, equal mobilities $\mu^-\!=\mu^+$, and $\tau_\text{e}=10$ ms ($\lambda_\text{e}/L=2.58$). 
The different curves show the absolute values of $E_x(x)$, $\rho^+(x)$, $\rho^-(x)$ and $\rho^\text e (x)\times 10^5$
 vs. $x/L$,  in units of $E_\circ$ and $\rho_\circ$. 
 }
\label{fig:1D-abso}
\end{center}
\end{figure}

The continuity equation for the average electron charge density, 
 in steady conditions, can be written as 
\begin{equation} 
\frac{ d (\rho^\text{e} v^\text{e}_x) } {dx}  = -K - \frac{\rho^\text{e}}{\tau_\text{e}}
\; ,  \label{eq:continuity-e}
\end{equation}
where $-K$ this time defines the injection of negative charge density, 
 drifting towards $x=0$ (with $\rho^\text{e}<0$,  $v^\text{e}_x<0$, $K>0$).
 An average capture length $\lambda_\text{e}= |v^\text{e}| \tau_\text{e}$ can be defined, and compared to the gap lenght $L$.
 
 The electron attachment is source of negative ions, which drift towards the anode with a mobility similar to the one of negative ions \cite{ion-mobility-2}, and contribute to the space charge present in the detector. 
The result of a numerical solution~\cite{Comsol}
to the continuity equations for $\rho^e,\, \rho^+,\, \rho^-$ and to the Gauss's low for $E_x$ 
 is shown in figure~\ref{fig:1D-abso} in the case of $\lambda_\text{e}(E_\circ)/L=2.58$,
$\alpha=1.15$, and assuming equal mobilities for negative and positive ions. 
The presence of the negative ions causes the minimum of the electric field to move away from the anode to a shallow minimum 
 at $x_\text{min}/L \simeq 0.15$ , where the field is 3\% lower than at the anode; in comparison with the case of infinite lifetime, the field at the anode is increased from $0.77\,E_\circ$ to $0.83\,E_\circ$, and the field at the cathode is barely changed from $1.38\,E_\circ$ to $1.37\,E_\circ$. 
For an electron lifetime shorter by a factor 2 (i.e.\ $\lambda_\text{e}/L=1.29$), the corresponding value are $x_\text{min}/L\simeq 0.24$, $E_\text{min}/E_\text{a} = 0.94$, $E_\text{a}/E_\circ=0.90$, $E_\text{c}/E_\circ= 1.35$.

An approximate analytical expression for $E_x(x)$, which matches the numerical evaluation at the level of 1\%, is provided in appendix \ref{app:1D-lifetime}.

In the following sections, analytical results are obtained ignoring the dependence of the initial recombination 
on the electric field, while the effect is included in numerical computations.


\section{Dual-phase detectors and feedback of positive ions}  \label{sec:dualphase}
Dual-phase scintillation and ionisation detectors have been used for the study of rare processes, and large devices based on argon have been proposed for rare processes and neutrino experiments~\cite{interface-block,dual-phase-WA105}.
After drifting in the liquid, the electrons from primary ionisation are extracted into the vapour phase,
where they can be accelerated in order to produce a light signal or to achieve charge amplification~\cite{dual-phase-gain}.
In the latter case, 
a significant fraction of positive ions from the multiplication process may be drawn into 
the liquid, contributing to the 
space charge in the drift volume.

Near the liquid-vapour interface, an extraction grid is designed to establish a higher electric field strength, which facilitates 
the transition of the electrons from the liquid to the vapour. 
 At the interface, dielectric polarisation attracts  charges in the vapour phase -- of any sign -- towards the interface surface, 
and repels them from the surface when they are in the liquid (see for example \cite{Jackson}). 
The higher field strength obtained with the extraction grid 
is designed to counteract this effect, together with any effective binding potential for conduction electrons in the liquid phase~\cite{binding-potential}.  It has been argued that the 
discontinuity of the polarisation field at the interface may prevent positive ions from entering the liquid~\cite{interface-block}.
Opposite arguments have also been presented~\cite{interface-favor}. 
Lacking direct evidence, it is assumed here that positive ions can reach and cross the vapour to liquid interface. Polarisation and binding potential effects appear more likely to play a role in preventing negative ions, which may come 
from electron attachment, from leaving the liquid phase~\cite{negative-ion-block}, as they do for electrons.
That would be a minor effect under the assumption  $\lambda_\text{e} \gg L$ and, furthermore, a build-up of negative ions on the liquid-vapour interface would affect more directly the electron extraction from the liquid phase, 
rather than the electron drift in the liquid TPC.

The relative amount of ions feedback to the drift region is described by the product  $\beta = (g-1) \times f$, 
where $g$  is the gain on the electron signal and the factor $f$ includes the collection of the positive ions from the amplification region, their transfer into the liquid phase, and the limited transparency of the extraction grid for positive charges. 
The value of $\beta$ depends on the electric field at the grid, on the side of the drift volume, which corresponds to the
the field at the anode $E_\text{a}$ of single-phase detectors, and is affected by space charge. 
\begin{figure}[tb]
\begin{center}
\includegraphics[width=0.7\textwidth]{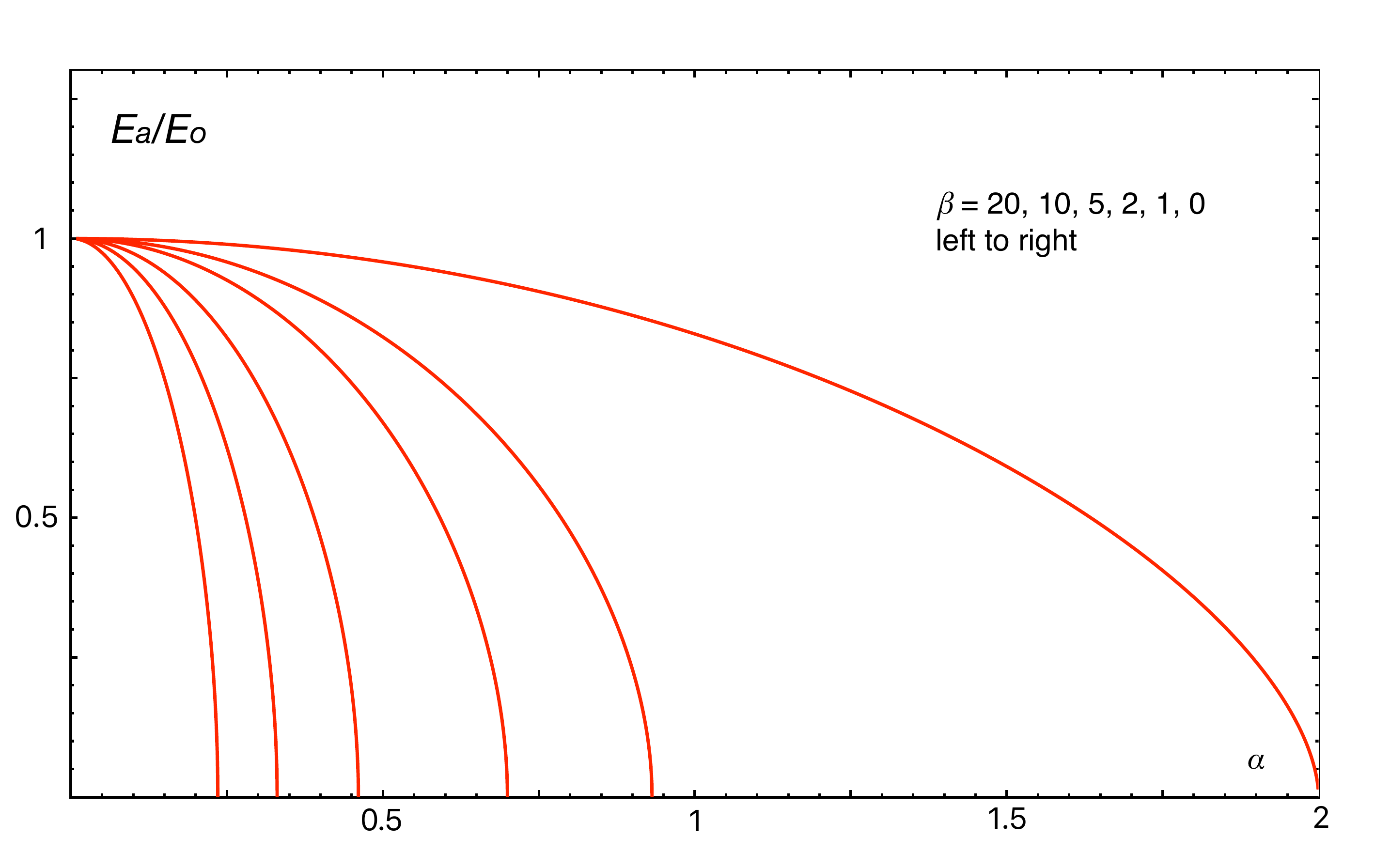}
\caption {Behaviour of $E_\text a/E_\circ$ vs.\ $\alpha$, for different values of the feedback parameter $\beta$.
}
\label{fig:backflow1Dcritical}
\end{center}
\end{figure}
\begin{figure}[t]
\begin{center}
\includegraphics[width=0.7\textwidth]{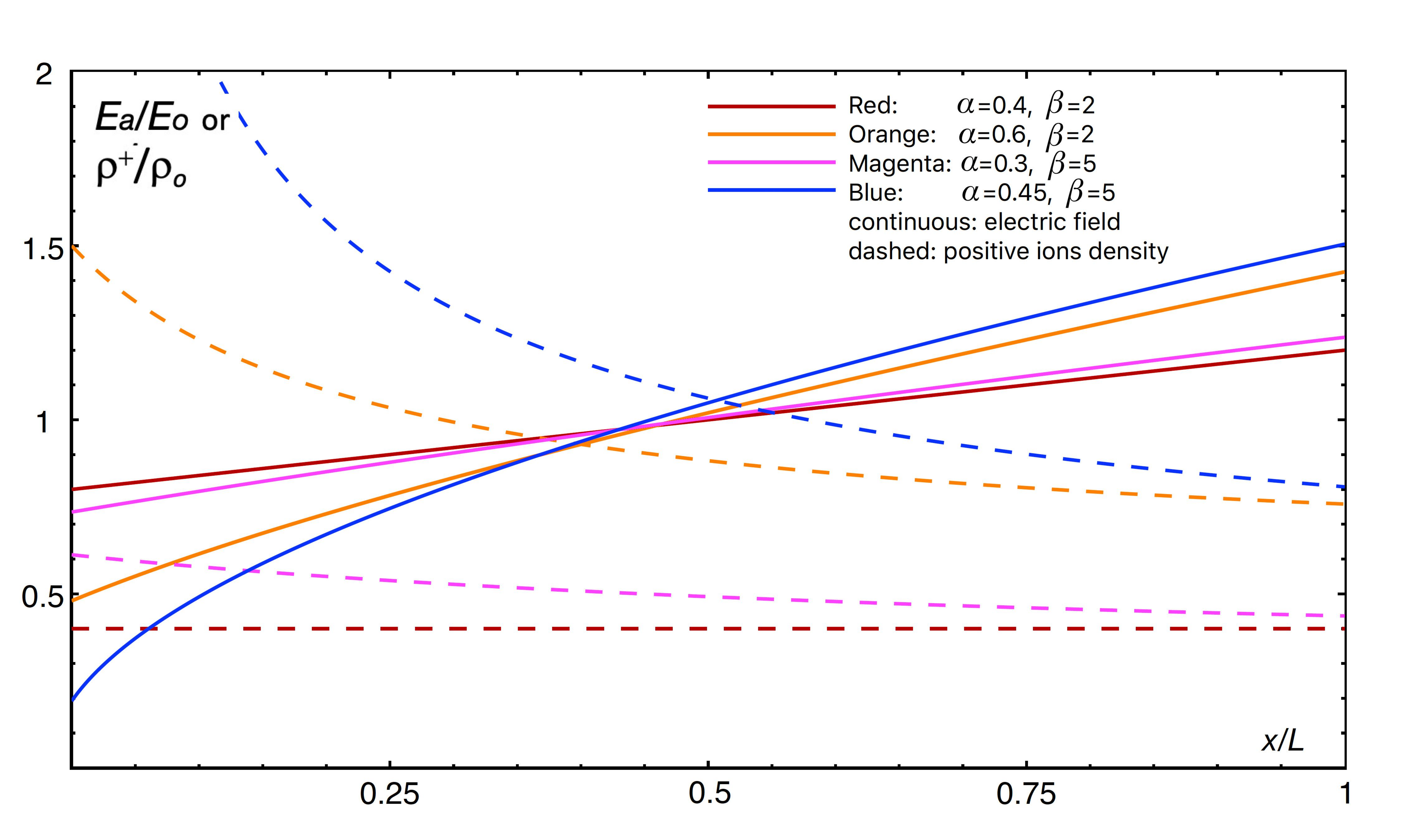}
\caption {Behaviour of $E_x(x)/E_\circ$ (continuous lines) and $\rho^+\! (x)/\rho_\circ$ 
(dashed lines) vs. $x/L$ for different values of the parameters $\alpha$ 
and $\beta$.}
\label{fig:backflow-profiles}
\end{center}
\end{figure} 
 
In steady conditions, and neglecting electron capture for the moment, the flux of electrons leaving the liquid phase is equal to the total rate of primary ionisation in the detector, $J_\text{e}=KL$, and the 
corresponding flux of positive ions crossing the extraction grid is 
\begin{equation}
J(x\!=\!0) = \beta \, K\, L \;.
\end{equation}
The differential equation describing the space charge in steady conditions is the same as eq.~\ref{eq:continuity}, but now the 
boundary condition includes $J(0)$ and the solution is 
\begin{equation}
\mu^+ E_x \, \rho^+\!(x) = K(x+\beta \,L) \; . 
\end{equation}
The differential equation describing the electric field can be be easily integrated also in this case, obtaining
\begin{equation}
E_x(x) =  E_\circ \sqrt{(E_\text a/E_\circ)^2 + \alpha^2 [(x/L)^2 + 2\, \beta \,(x/L)]} \;.
\end{equation}
The effect of space charge are now controlled by the two dimensionless parameters $\alpha$ and $\beta$. 
The reduction in $E_\text a$ as a function of $\alpha$ is significantly enhanced by the presence of feedback, 
as shown in figure~\ref{fig:backflow1Dcritical}. 
For $\beta \ge 1$, the critical condition of vanishing $E_\text a$  is reached for values $\alpha < 1$.
Figure~\ref{fig:backflow-profiles} shows the behaviour of $E_x/E_\circ$ and $\rho^+\!(x)/\rho_\circ$ 
for some values of the parameters $\alpha$ and $\beta$. While for $\beta=0$ the space-charge density $\rho^+(x)$ 
increases with $x$, the trend may be inverted when ion feedback is present, as shown in these examples. 

For detectors with large $L/E$ 
and feedback of positive ions, 
natural radioactivity alone may cause non-negligible effects of space charge.
With $^{39}\!$Ar radioactivity of about 1.0~Bq/kg in atmospheric argon, $<\!E_\beta\!>=220$~keV, 
effective ionisation potential 24~eV, and ionisation yield of 0.7 at 500~V/cm,  
the charge injection rate is $K=1.4\times10^{-12}$~C\,m$^{-3}$s$^{-1}$. 
With $L/E=$ 10 m / 500 V\,cm$^{-1}$ 
and $\mu^+=1.6\times 10^{-7}$~m$^2$V$^{-1}$s$^{-1}$, 
the value $\alpha=0.16$ is obtained, which implies significant (large) effects of space charge  for 
an ion feedback parameter $\beta > 2$ ($\beta > 20$),
as shown in figure~\ref{fig:backflow1Dcritical}.


\section{Mitigation technique n.\ 1\,: separation grid}  \label{sec:grid}
At first order in $\alpha^2$, eq.~\ref{eq:gauss} can be written as 
$dE_x/dx \simeq (K\, x)/(\epsilon\, \mu^+E_\circ)$, 
so that the difference in field strength between cathode and anode is approximately equal to   
$(K\, L^2)/(2\, \epsilon \,\mu^+ E_\circ) \simeq \alpha^2 E_\circ /2$.
This range of variation in field strength can be reduced by means of a third electrode, a grid placed at the coordinate $x_\text g$ that constrains $V(x_\text{g})$ to a suitable value $V_\text{g}$. 
The grid generates a discontinuity in $E_x$ that can be exploited to increase the field strength at the anode and decrease it at the cathode. 
With a higher field on the anode side of the grid, drifting electrons will cross it, while a fraction of positive ions are captured \cite{Blum-Rolandi}, 
reducing further the effects of space charge in the region between the grid and the cathode.

\begin{figure}[tb]
\begin{center}
\includegraphics[width=0.7\textwidth]{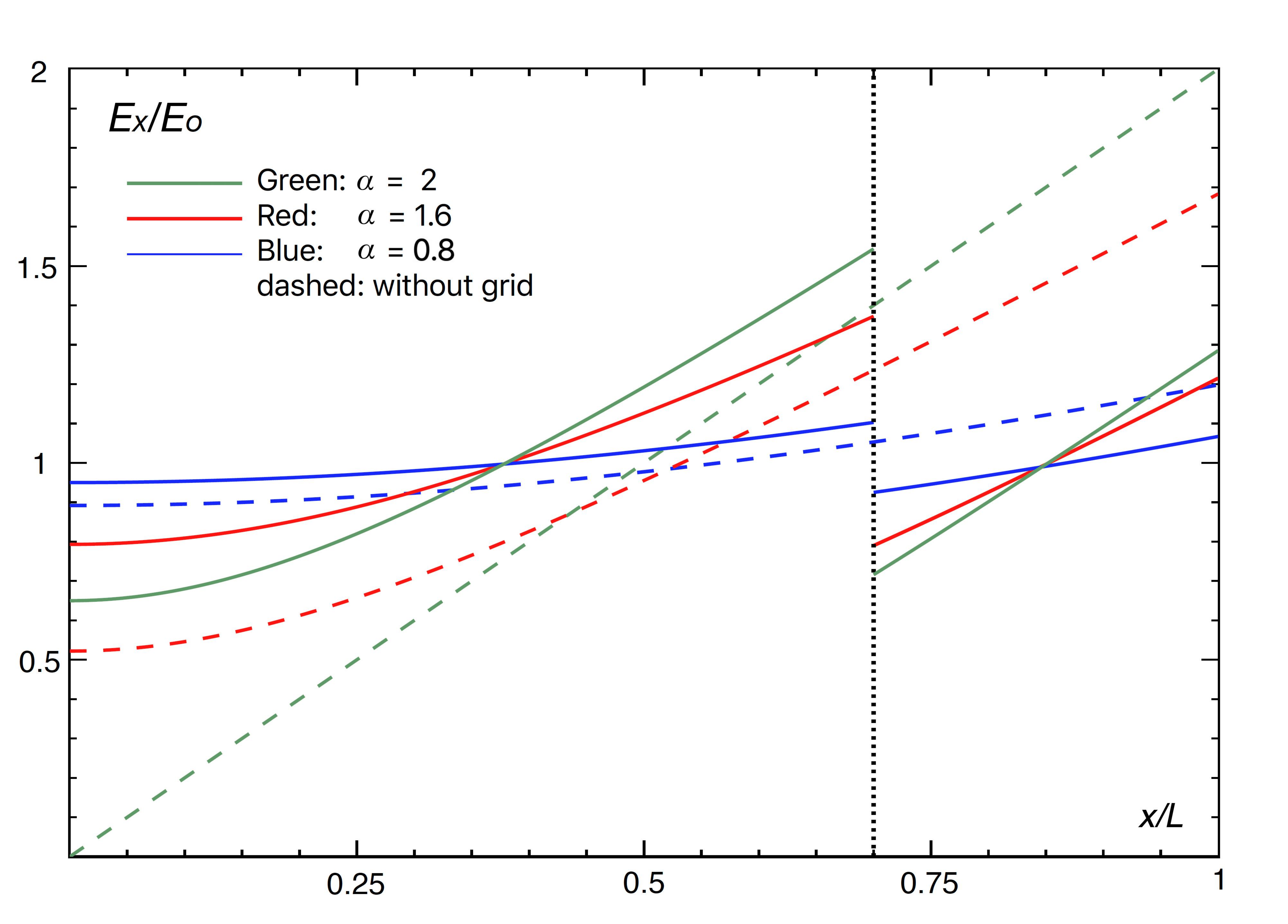}  
\caption {
Behaviour of $E_x(x)/E_\circ$ with a voltage grid placed at $x=0.7 \,L$, with $\alpha=0.8$ (blue), $1.6$ (red), and $2.0$ (green). 
The voltage at the grid is set to the value corresponding to a condition of no space charge ($\delta_\text g = x_\text g /L$). For comparison, the dashed lines show the electric field without voltage grid. 
} 
\label{fig:grid}
\end{center}
\end{figure}

\begin{figure}[h!]
\begin{center}
\includegraphics[width=0.7\textwidth]{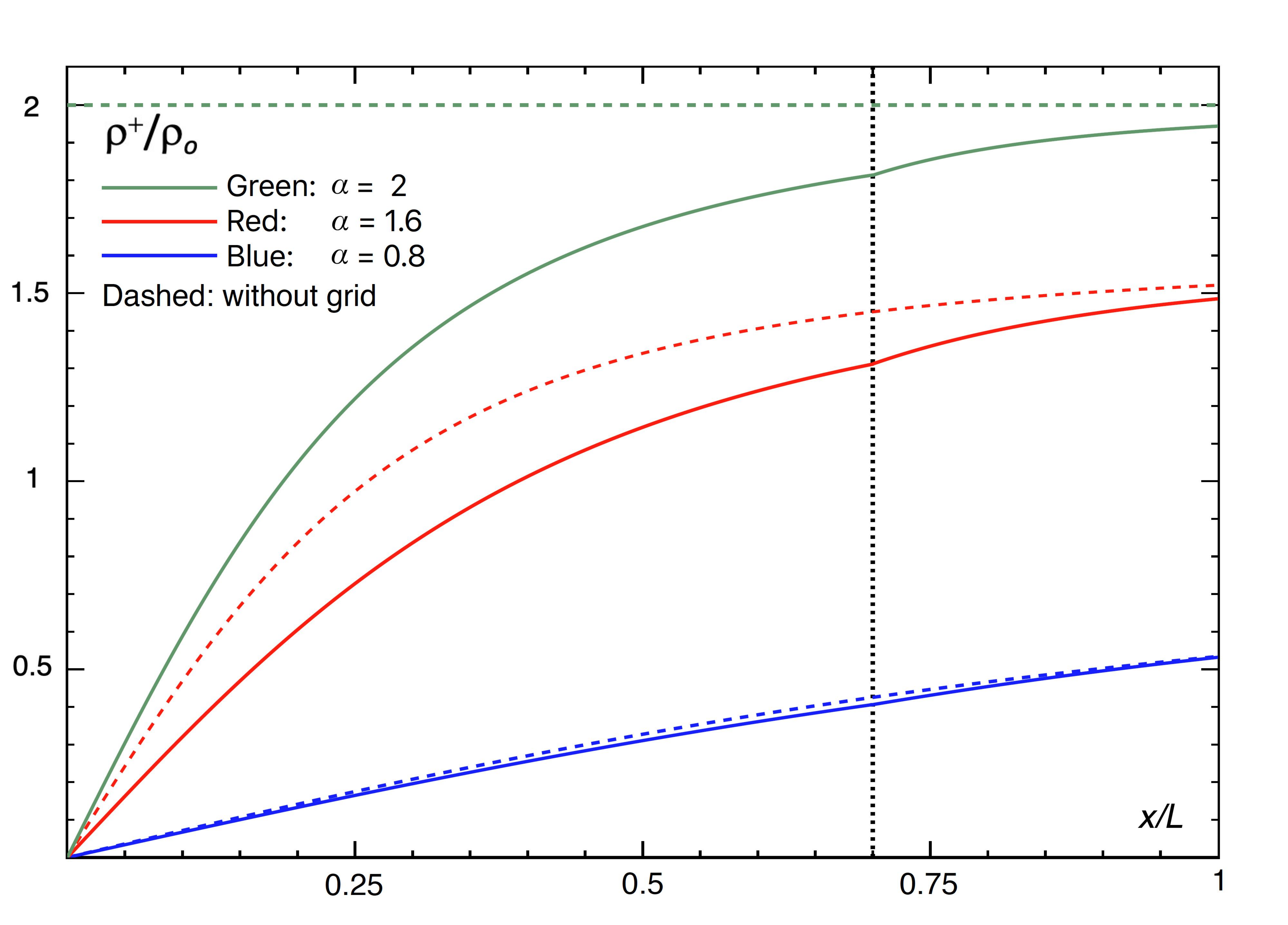}  
\caption {
Behaviour of $\rho^+\! (x)/\rho_\circ$ with a voltage grid placed at $x=0.7 \,L$, with $\alpha=0.8$ (blue), $1.6$ (red), and $2.0$ (green).
The voltage at the grid is set to the value corresponding to a condition of no space charge ($\delta_\text g = x_\text g /L$). For comparison, the dashed lines show the space-charge density without voltage grid. 
} 
\label{fig:grid-rho}
\end{center}
\end{figure}

Figure \ref{fig:grid} illustrates the value of $E_x(x)/E_\circ$ for a given configuration of $x_\text{g}$ and $V(x_\text{g})$, and different conditions of space charge, corresponding to $\alpha$ 
equal to 0.8, 1.6 and 2, and no feedback of positive ions ($\beta = 0$).
The grid reduces by at least a factor 2 the range between the highest and the lowest values of the field strength across the full gap.
Figure \ref{fig:grid-rho} shows the corresponding distribution of $\rho^+(x)/(\rho_\circ)$.
For $x<x_\text{g}$, the electric field is described by eq.~\ref{eq:Ex} as in the case without grid, 
but now the boundary condition on the field at the anode $E_\text{a}$ is
\begin{equation} \label{eq:Ea-grid}
\int_0^{x_\text{g}} \sqrt{(E_\text{a}/E_\circ)^2 +(\alpha \, x / L)^2} \; dx= \delta_\text{g} \, V_\circ \;,
\end{equation}
with $\delta_\text{g} = -V_\text{g}/V_\circ$ and, as usual, $V=-V_\circ$ (0) at the cathode (anode). 
The boundary condition can be written as 
\begin{equation}
 \int_0^1 \sqrt{\left(\frac {E_\text{a}\, x_\text{g}}{E_\circ \delta_\text{g}\,L}\right)^{\!2} +
\left( \frac{\alpha \, x_\text{g}{}^2\, x^\prime}{\delta_\text{g}\,L^3} \right)^{\!2} }\; dx^\prime = L \; ,
\end{equation}
with $x^\prime = x\times L/x_\text g$.

As discussed below, a convenient configuration is with $x_\text{g} /L = \delta_\text{g}$, in which the grid restores at $x_\text g$ the voltage that would be obtained without charge injection. Then the boundary condition becomes 
\begin{equation}
\int_0^1 \sqrt{(E_\text{a}/E_\circ)^2 +(\alpha \, x_\text{g} \, x^\prime/ L^2)^2} \; dx^\prime= L \;,  \label{eq:grid-boundary-anode}
\end{equation}
showing that $E_\text{a}$, together with $E_x(x)$ for $0 < x < x_\text g$,
reproduces the solution for $0 <  x < L$ obtained in the case without grid and with $\alpha$ replaced by $\alpha \times x_\text{g}/L$.
Therefore, the distortion to the values of the electric field are reduced by a factor equal to the square of  $x_\text{g}$, 
and the longitudinal distortion in a TPC device by the cube of $x_\text{g}$.

As shown in appendix \ref{app:1D-grid}, for $x > x_\text{g}$, the general solution for the $E_x(x)$ is 
\begin{equation}
E_x(x) = E_\circ \sqrt{(E_{\text {g}+}/E_\circ )^2 +\alpha^2[(x-x_\text g)^2/L^2
+2\, (E_{\text {g}+}/E_{\text{g}-}) \, x_\text g\, (x-x_\text g)/L^2]} \hspace{0.5cm}  (x > x_\text{g}) \;,     \label{eq:grid-E-cathode}
\end{equation}
where $E_{\text {g}-}$ and $E_{\text {g}+}$ are the field values at the grid, on the sides of $x<x_\text g$, $x>x_\text g$ respectively, and are determined by the values of $x_\text g$ and $\delta_\text g$, together with those of $E_\circ$ and $\alpha$. 
The flux of positive ions $\rho^+\!v^+$ is reduced by the factor $E_{\text {g}+}\!/E_{\text{g}-}$ as it crosses the grid. The same factor applies for the change in ion drift velocity, so that derivative $dE_x/dx$ and the density $\rho^+$ are continuous across the grid, as shown in figures~\ref{fig:grid} and \ref{fig:grid-rho}.

The parameters $x_\text g$ and $\delta_\text g$ can be chosen so that in the two regions defined by the grid, the ranges in the value of  $E_x$ 
are approximately equal.  This is approximately equivalent to aiming at the maximisation of  the lowest value of $E_x$ across the full gap.  
As discussed in more detail in appendix \ref{app:1D-grid}, choosing $\delta_\text g = x_\text g /L$, i.e. restoring at $x_\text g$ the 
voltage that would be obtained without space charge and without grid, approximates the optimal configuration. 
The optimal value of $x_\text g/L$ is in the approximate range 0.6--0.7, with limited sensitivity to the exact value, with a 
preference for the lower (higher) values for $\alpha$ larger (smaller) than 1. 

As shown in figure~\ref{fig:grid}, the grid changes the range of  $E_x - E_\circ$ by a factor approximately equal to $(x_\text g/L)^2 \simeq 0.5$. 
The effect is larger when critical conditions are approached:  for  $\alpha = 2$, $\beta = 0$,  the grid reduces the electric field variation from $\pm 100$\% to about $\pm 30$\%.

As a final remark, the separation grid is charged with negative surface charge density 
$\sigma_\text g = - \varepsilon (E_{\text g -}\!-E_{\text g +})$
and subject to the electrostatic pressure  
$p_\text g = - \varepsilon (E_{\text g -}{}^2 \! -E_{\text g +}{}^2)/2$. 
To first order in $\alpha^2$, the pressure is given by the expression  $p_\text g \simeq - 0.24\, \varepsilon \, E_\circ^2 \, \alpha^2 $, 
which is accurate within 5\% for the examples shown in figures~\ref{fig:grid} and \ref{fig:grid-rho}.
In practical situations it may be similar to the pressure on the anode 
$(p_\text a \simeq 0.5 \, \varepsilon \, E_\circ^2 \, (1-\alpha^2 /3))$, 
and is smaller than the pressure on the cathode 
$(p_\text c \simeq -0.5\,\varepsilon \, E_\circ^2 \, (1+2\,\alpha^2/3))$.

\section{Side walls, field cage and transverse effects}   \label{sec:sidewalls}
Besides the effect on the component of the electric field driving the drift of electrons and ions discussed above, 
which can be referred to as \emph {longitudinal},
in practical conditions the space charge can induce \emph {transverse} distortion. 
The main reason for this effect is that liquid-argon time-projection chambers~\cite{Mooney:2015kke, Antonello:2020qht}
have been operated with side walls equipped with field cages designed for the operation without space charge, 
i.e.\ with a uniform gradient of the voltage $V_\text{fc}$ established by the field cage ($dV_\text{fc}/dx=-E_\circ$). 
This field pattern does not match the field established far inside from the field cage, 
and a transverse component of the electric field arises close the side walls.

\begin{figure}[t]
\begin{center}
\includegraphics[width=0.6\textwidth]{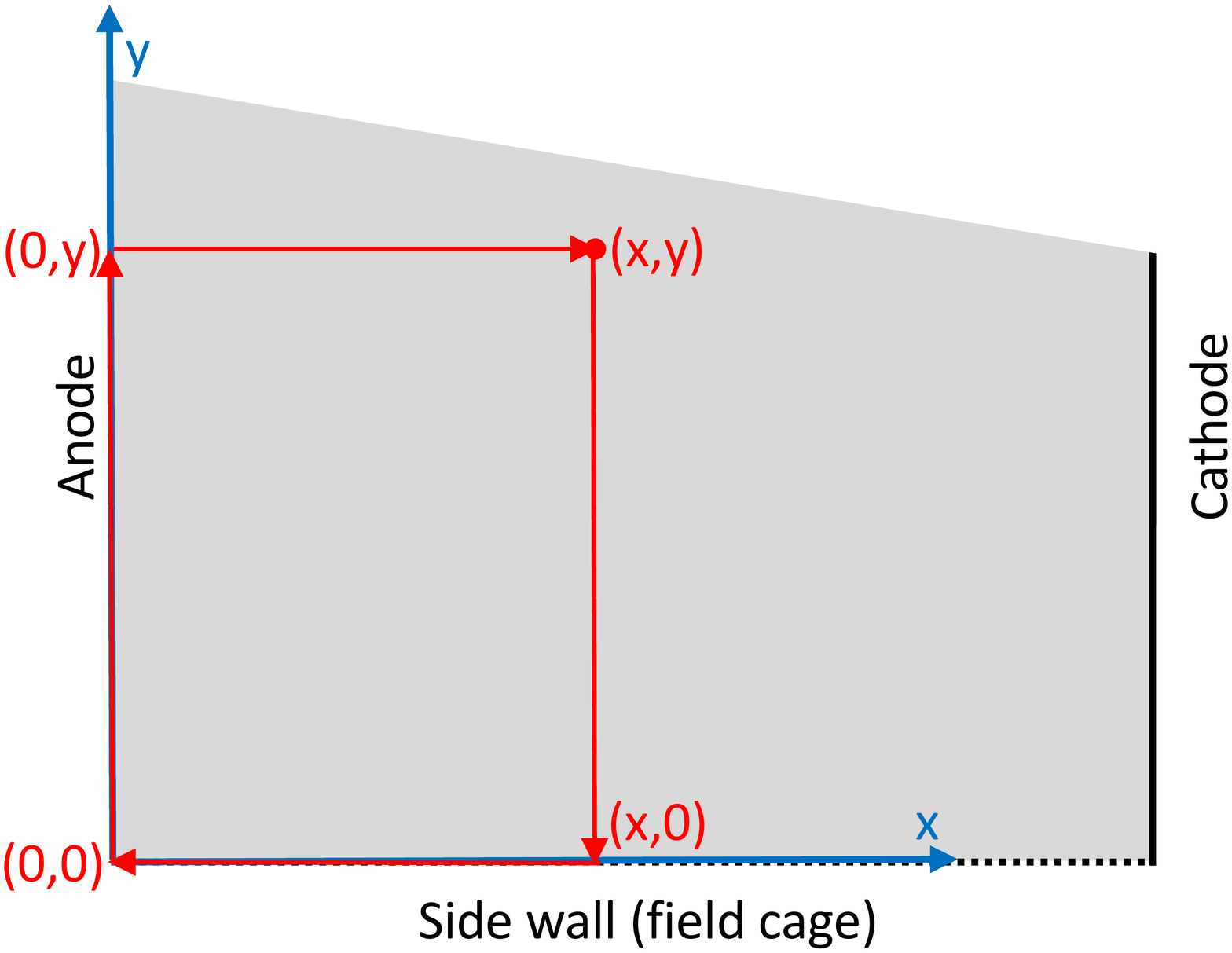} 
\caption {Path for line integral of electric field, as discussed in the text.} 
\label{fig:contour-plot}
\end{center}
\end{figure}

A constraint on the transverse components of the electric field  $E_y$, $E_z$ can be placed considering line-integrals
$\oint \overline {\vphantom{d} E} \, \overline {ds}$ computed along closed paths.  
For the component along $y$, consider the path of four straight segments shown in figure~\ref{fig:contour-plot}, 
which starts at a point ($x, y, z$) far from the field cage, 
(a) reaches the field cage at $(x, 0, z)$, 
(b) reaches the anode at ($0, 0, z)$, 
(c) follows the anode to ($0, y, z$),
and (d) closes the path to ($x, y, z$).  
The contribution from (c) is null, so that the opposite of the term in (a) is equal to the sum of the term in (b) and (d), and  
satisfies on the line--integral condition 
\begin{eqnarray}
\int_{(x,0,z)}^{(x,y,z)} E_y(x,y'\!,z)\,dy' & = & \int_{(x,0,z)}^{(0,0,z)} E_x(x'\!,0,z)\,dx' + \int_{(0,y,z)}^{(x,y,z)} E_x(x'\!,y,z)\,dx' \nonumber \\
                                       & = & -E_\circ\, x - V(x,y,z)    \label{eq:Et-int}\\    
                                      &\simeq & -E_\circ\, x -V(x)  \nonumber
\; ,
\end{eqnarray}
where the electric voltage at the field cage is $-E_\circ\,x$, 
the dependences of $V(x,y,z)$ on $y,\, z$ can be neglected because of the distance from the field cage,   
and $V(x)$ is the electric voltage in the one-dimensional description of space-charge effects 
obtained from the integration of $-E_x(x)$ in eq.~\ref{eq:Ex}. 
Since for $0 < x < L$ the absolute value of  $V(x)$ is smaller than $E_\circ\,x$, 
the transverse electric field is negative (directed towards the field cage), and the drifting electron are focussed toward the center of the detector. 
The value of $V(x)$ and and its approximation to first order in $\alpha^2$  are provided in appendix~\ref{app:analytical expressions}. 
For $\alpha < 1.5$, the absolute value of the integral is largest 
at $x \simeq L/\sqrt{3}$ and is approximately equal to  $0.064\,\alpha^2 E_\circ \, L$. For $\alpha$ approaching 2, the maximum is moved towards $x \simeq L/2$ with the value $0.25\, E_\circ \, L$.

\begin{figure}[tb]
\begin{center}
\includegraphics[width=0.89\textwidth]{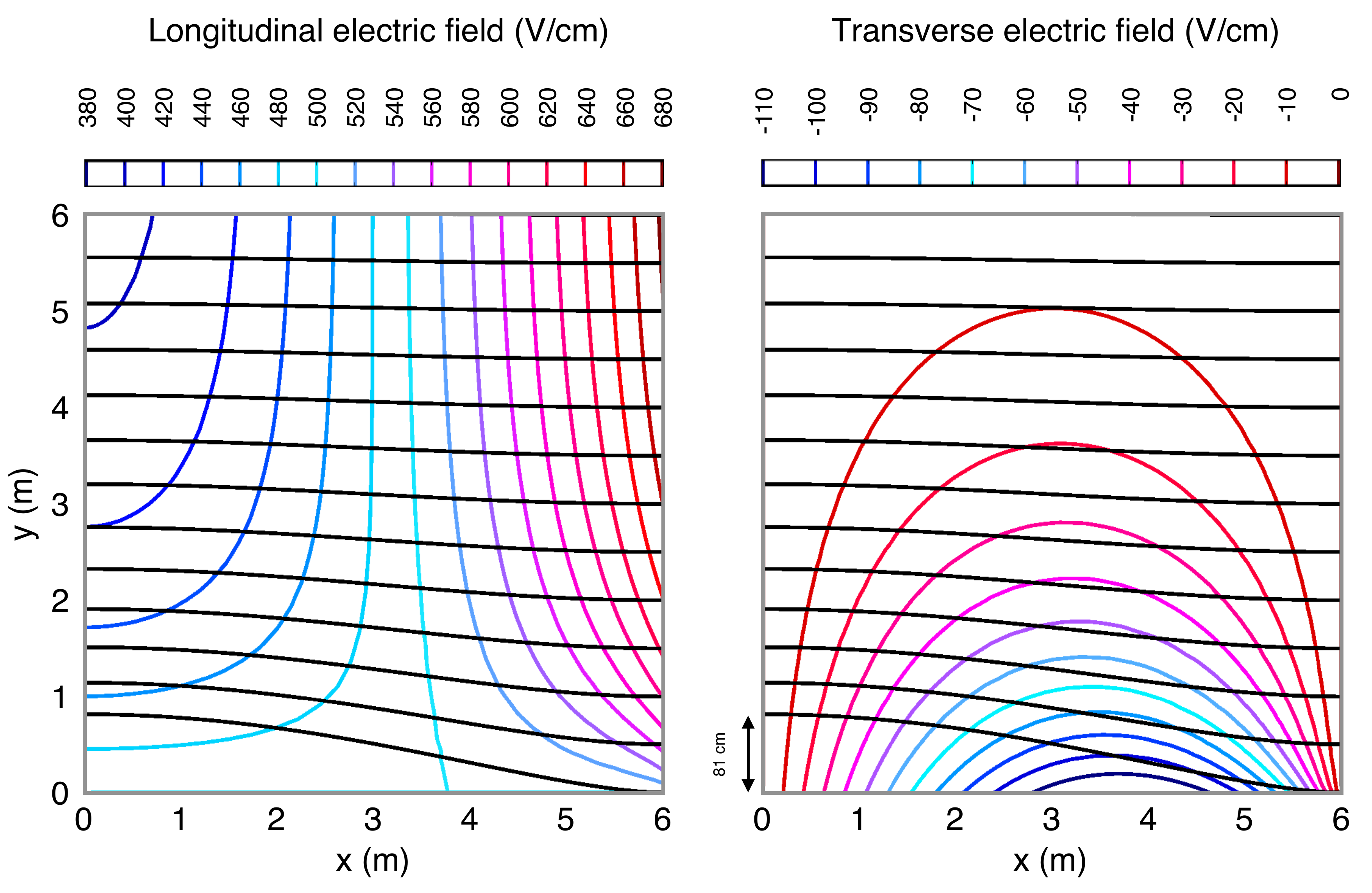}  
\caption {
Contours of equal values of the longitudinal $E_x$ \emph {(left)} and transverse $E_y$ \emph {(right)} components of the electric field from a numerical calculation.
The plots cover the anode to cathode region, 6~m long, and a 6~m wide region with the field cage at $y=0$. The computation neglects dependences on the thirds coordinate $z$, and assumes a detector transverse size of 20 m along $y$. 
Input values are $E_\circ = 500$ V/cm and $\alpha=1.15$. The contours cover the range of $E_x$ from 
390 to 670~V/cm, and of $E_y$ from 10 to 110~V/cm. The black line shows the drift path of electrons from the cathode to the anode.}
\label{fig:2DExEy}
\end{center}
\end{figure}

The solution of a set of dependent differential equations describing the electric field and the motion of charge species
can be numerically approximated using FEA software~\cite{Comsol}. Computations have been made
in three dimensions, 
or in two dimensions 
when the third coordinate is far from the field cage. 
Figure~\ref{fig:2DExEy} shows an example of contours of equal values of $E_x$ and $E_y$ in a two dimensional computation. 
If the transverse size of the detector is large compared to the anode to cathode distance $L$, the fundamental parameters that define the electric field are the same as in the one-dimensional case:  
$E_\circ$, $L$, and $\alpha^2$, with the latter proportional to $(L/E_\circ)^2$.
Therefore, $(E_x-E_\circ)/E_\circ$ and $E_y/E_\circ$ scale as $\alpha^2$, and $L^2/E_\circ^2$, 
spatial distortions $\delta x$, $\delta y$ in a TPC detector scale as $\alpha^2 L$, and $L^3/E_\circ^2$,
when all quantities are computed on coordinates that scale as $x/L$ and $y/L$. 
Numerical computations show that these scaling relations remain valid also for finite electron lifetime, if  $\lambda_\text{e}
\equiv  |v^\text{e}(E_\circ)| \tau_\text{e} \gg L$.

\begin{figure}[htb]
\begin{center}
\includegraphics[width=0.9\textwidth]{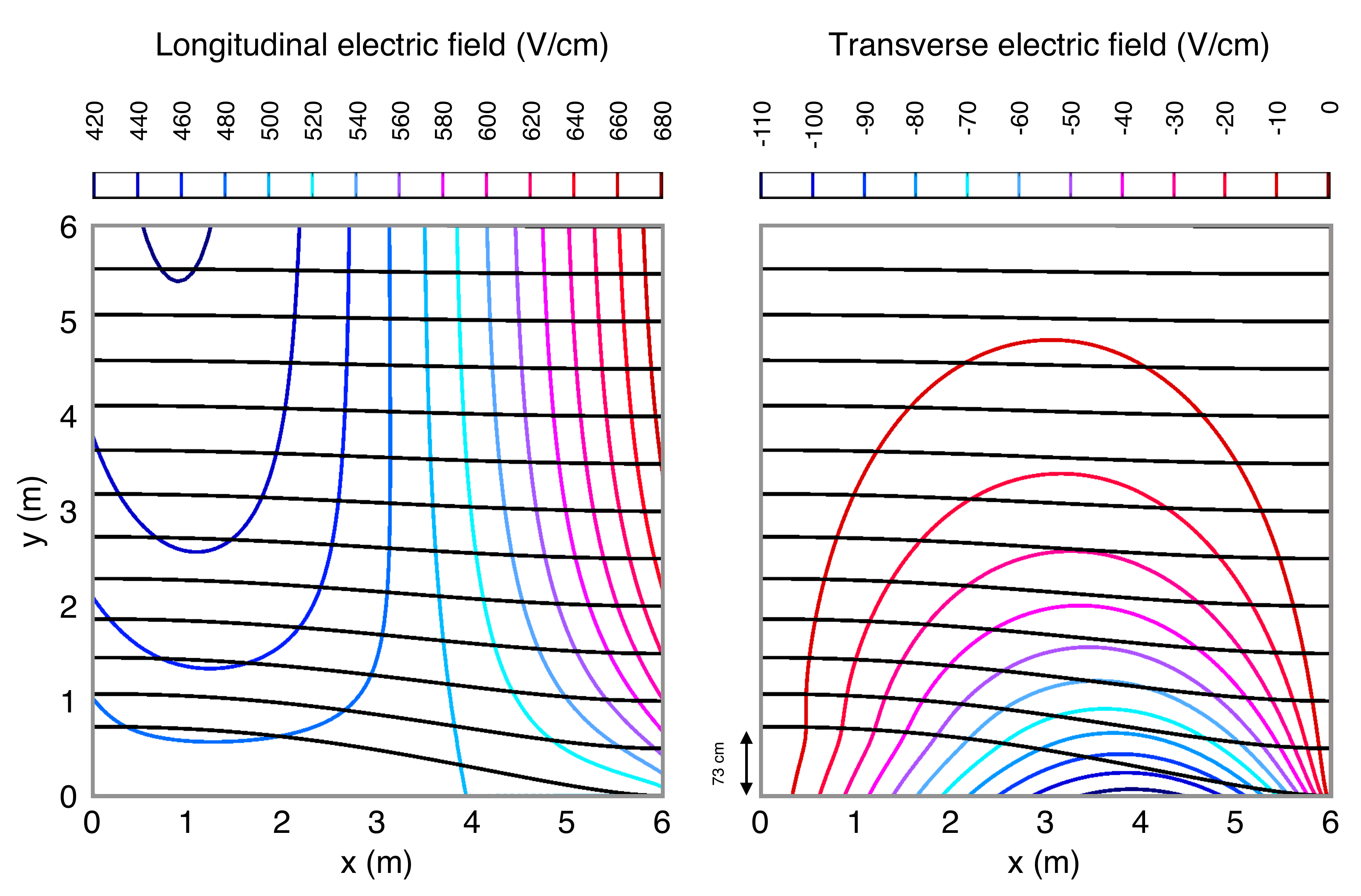}  
\caption {Same as in figure~\ref{fig:2DExEy}, but now with electron attachment corresponding to $\tau_\text{e}=10$ ms, and equal mobility for positive and negative ions.}
\label{fig:2DExEy-tau}
\end{center}
\end{figure}
On the side wall ($y=0$), i.e.\ on the field cage, the largest value of of the transverse component $|E_y|$ occurs for 
$x \simeq 0.63\,L$, and is equal to about $0.18\, \alpha^2 E_\circ$.
Equation~\ref{eq:Eymax} in appendix~\ref{app:analytical expressions} provides a polynomial approximation of the 
transverse component of the electric field at the field cage ($E_y(x,0)$). The transverse component  
decreases with $y/L$, following approximately the exponential dependence 
$E_y(x,0)\times \exp [-y/\lambda_y(x)]$,
with $\lambda_y(x)\simeq 0.47\,[1-0.5(x/L-0.5)]\,L $.\footnote  
{This value is obtained from the ratio $\lambda_y = [V(x)-E_\circ x]/E_y(x,0)$, 
with $V(x)$ from the one-dimensional model and $E_y(x,0)$ from the polynomial fit.   
A better approximation to $E_y(x,y)$ for $x \simeq L/2$ is provided in eq.~\ref{eq:EyL2y}.
}
At a distance from the side wall equal to 0.5 (1.0) $L$, the maximum value of the transverse field occurs at $x\simeq L/2$
and is equal to about 4.1\% (1.1\%) of $\alpha^2 E_\circ$. 

At the field cage, the longitudinal component $E_x(x,0)$ is constrained to $E_\circ$. 
For $y>L/2$ and $x \simeq L/2$, $E_x(x,y)$ is described by the one-dimensional approximation (eq.~\ref{eq:Ex}) 
within  about 1\%.  Near the electrodes, the one-dimensional approximation is valid for $y > L$ to the same accuracy.

A finite electron lifetime, 
if significantly larger than the electron maximum drift time, does not alter the scenario, as shown in the comparison of 
figures~\ref{fig:2DExEy} and \ref{fig:2DExEy-tau}.  The latter is obtained from a numerical calculation 
using $\lambda^\text{e}_\circ /L = 2.58$.  
As discussed above in section~\ref{sec:electron-lifetime}, the presence of negative ions reduces the distortion in $E_x$
(the range $-$22\% to $+$38\% is reduced to $-$17\% to $+$34\% for $\alpha=1.15$, with the minimum of $E_x$ moved to $x\simeq 0.15$). 
The maximum values of $|E_y|$ is reduced by about 10\%. The lateral extension of the region with significant transverse field is not significantly modified. 

The ionisation electrons drift following the direction of $-\overline E$, and the transverse component $E_y$ 
causes a displacement of the point of collection of the electrons starting at  $(x, y, z)$ according to the expression
\begin{equation} 
\delta y(x,y,z) = \int_{(x,y,z)}^{x=0} \frac{E_y}{E_x}dx \; ,
\end{equation}
with the integration computed along the path followed by the drifting electrons.
Under usual conditions, this \emph {transverse distortion\/} collects same-sign contributions along the full drift path, and for an initial coordinate $x > L/2$, 
its value can be significantly larger than the longitudinal distortion of 
eq.~\ref{eq:longitudinal-dist}, because the latter is reduced by competing contributions of different sign from  the regions 
$x\gtrsim L/2$ and $x \lesssim L/2$,
and also the reduced dependence of the electron drift velocity on the electric field  ($\gamma = (\delta v\,^e / v\,^e)/(\delta E_x /E_\circ) \simeq 0.5$ for $E_\circ \simeq 500$~V/cm). 
The largest transverse distortion occurs for drift paths starting at $x=L$ and $y=0$, where the numerical computation 
provides $\delta y_\text{max} = \delta y(L,0,z) = 0.105 \, \alpha^2 L$, which is three times larger than the corresponding maximum longitudinal distortion 
(the coordinate $z$ is assumed here to be far from the corresponding side walls). 
%
Drift paths are shown as black lines in figures~\ref{fig:2DExEy} and \ref{fig:2DExEy-tau}.

The transverse distortion $\delta y(x,y,z)$ results in a scale distortion along the $y$ direction equal to $d (\delta y)/dy$,
which should be taken into account for measurements of specific signal yield $dQ/dX$, or when establishing 
intervals on the $y$ coordinate. 
The largest distortion to the length scale $dX$ occurs at the cathode and is approximately equal to  
$- 0.2 \, \alpha^2 \exp(-2 \,y/L) \cos^2 \theta_y$, where $\theta_y$ is the angle between the direction of $dX$ and the $y$ axis.
%
%
 \section{Detector aspect ratio}  \label{sec:aspect-ratio}
If the transverse size of the drift volume (with widths $W_y$, $W_z$) is not much larger than the drift gap $L$, the one-dimensional description of sections~\ref{sec:1D-model}--\ref{sec:grid} needs to be revised. General features and numerical examples are presented in this section. 

As discussed in section~\ref{sec:sidewalls}, with $L \ll W_z$ and for the $z$ coordinate far from the edges $z=0$, $W_z$, the transverse component of the electric field decreases  as  
$E_y(x,y)\approx  E_y(x,0) \times \exp(-y/\lambda_y(x))$ as the distance $y$ from the field cage increases, with $\lambda_y \simeq L/2$. 
If $W_y$ is not much larger than $L$, there is not enough width to reach a negligible value of $E_y$ before
approaching the centre of the drift volume $y \simeq W_y/2$.  
Because of symmetry with the other half of the detector, $E_y$ still vanishes at y=0, but does it 
with a finite gradient   
\begin{equation}
E_y(x,y) \simeq  \kappa (x)\times (y-W_y/2)\hspace{1cm} (y\simeq W_y/2) \,  .
\end{equation}
The function $\kappa(x)$ vanishes at the electrodes  ($x=0$, $x=L$), and 
is a first order quantity in $\alpha^2$ in the range $0 < x < L$. 

Table~\ref{tab:transverse-gradient} shows the result of a numerical computation of $\partial E_y/\partial y$ for different detector 
aspect ratios, at coordinates  $x/L=0.5$, $y/W_y = 0.125$, 0.25, 0.5, 
for $\alpha =1.15$, $E_\circ = 500$~V/cm, $\tau_e=10$~ms, and equal mobility of positive and negative ions. 
The dependence on $z$ is ignored in the first three rows  ($W_z \gg L$ and $L < z < W_z - L$), while in the last row 
$W_z=L$ and the gradient is shown at $z=W_z/2$,  where $\partial E_z/\partial z = \partial E_y/\partial y$. 
\begin{table}[bt]
\begin{center}
\caption {Gradients of the transverse electric field  $\partial E_y/\partial y$  for different values of $y$ and different detector geometries, computed numerically at 
$x=3$~m, $z = W_z/2$, for $\alpha = 1.15$ and $\tau_e = 10$ ms. In the Gauss's equation for $\partial E_x/\partial x$, 
the transverse gradient is subtracted from the charge density 
$\rho /\epsilon \simeq 0.53$~V~cm$^{-2}$. In the last row, the gradient $\partial E_z/\partial z$ contributes an equal subtraction.
}
\label{tab:transverse-gradient}
\begin{tabular}{| c | c | c |}
\hline 
$\hphantom{ai}L \hphantom{.}\times \hphantom{i} W_y \hphantom{i} \times  \hphantom{.}  W_z $   &  y=3 m   	&  y= 6 m 	\\
\hline                                     
 6 m $ \times $ 24 m $\times \; \mbox {\Large $  \infty $}$       &  $\partial E_y/\partial y = 0.12$ V cm$^{-2}$ 	&  $\partial E_y/\partial y = 0.026 $ V cm$^{-2}$ \\
 6 m $ \times$ 12 m $ \times \; \mbox {\Large $  \infty $}$  &  $\partial E_y/\partial y = 0.12$ V cm$^{-2}$ 	&  $\partial E_y/\partial y = 0.052 $ V cm$^{-2}$ \\
 6 m $ \times \;$ 6 m $\;  \times \; \mbox {\Large $  \infty $}$    &  $\partial E_y/\partial y = 0.22$ V cm$^{-2}$ 	&    \\
 6 m $ \times \;$ 6 m $\; \times $ 6 m     &  $\partial E_y/\partial y = 0.17$ V cm$^{-2}$ 	&    \\
\hline 
\end{tabular}
\end{center}
\end{table}

Because of $\partial E_y/\partial y$, $\partial E_z/\partial z$, the Gauss' equation for the main component 
$\partial E_x/\partial x$ used in section~\ref{sec:1D-model} is modified.   
For a detector with $W_z=W_y=W$, near the axis $y = z =W/2$, both transverse component of the electric field contribute
equally:
\begin{equation} \label{eq:Gauss-kappa}
\frac{\partial E_x}{\partial x}=  \frac{\rho}{\epsilon}-\frac{\partial E_y}{\partial y} - \frac{\partial E_z}{\partial z}
\simeq  \frac{\rho }{\epsilon} -2 \,\kappa \hspace{1cm} (y \simeq  z \simeq W/2)\, \; . 
\end{equation}
Both $\rho/\varepsilon$ and $\kappa$ are positive and first order in $\alpha^2$, so that  $\kappa$,
which is due to the relative proximity of the linear field-cage, reduces the effects of space charge at the center of the
detector.\footnote
{The continuity equation for the space-charge density is also affected by the parameter $\kappa$, but less directly than $E_x$. Near the detector axis and considering positive ions, the transverse components of the current density satisfy  $\rho^+ v_y{}^+ \simeq \rho^+ \mu^+ \kappa \: (y-W_y/2)$, and similarly for $\rho^+ v_z{}^+$. The gradient of $\rho^+$ vanishes along the $y$, $z$ directions, 
so that the continuity equation 
becomes 
$\partial_x (\rho^+\! E_x) \simeq K/\mu^+ - 2\, \kappa\, \rho^+$.
The second term on the right reduces the amount of space charge stored in the detector, 
but it contributes at order $\alpha^4$, with limited 
effects unless $\alpha$ approaches the critical value. The same conclusion holds if negative ions are taken into consideration.}
The values shown in Table~\ref{tab:transverse-gradient} should  be compared to 
$\rho /\epsilon$, which is equal to about 0.53~V~cm$^{-2}$ at the center of the drift volume.  Using the Gauss' 
equation, the different configurations shown in the table correspond to a reduction of $\partial E_x/\partial x$ at the center of the drift volume, compared to the value obtained for $L \ll W_x, \, W_y$, in the range of 10 to 65\%. 

The values of $\partial E_y/\partial y$ shown in Table~\ref{tab:transverse-gradient} 
can be scaled to different intensities taking into account that they are proportional to $\alpha^2$. Furthermore, 
for $L=W_y \ll W_z$ or $L = W_y = W_z$,
and $\lambda_e \gg L$,   the gap $L$ is effectively the only parameter with the dimension of length, so that 
at the center of the detector:
\begin{eqnarray}
\kappa (L/2) = \: 0.20 \; \alpha^2 \, E_\circ / L \; ,& \hphantom{aaaa}  (L=W_y \ll W_z) \\
\kappa (L/2) = \: 0.15 \; \alpha^2 \, E_\circ / L \; .& \hphantom{aaaa}  (L=W_y =W_z) 
\end{eqnarray}
 In the latter case, $\partial E_y/\partial y$ is smaller because the amplitude of $E_y(x,y,z)$ is reduced by the constraints
 $E_y(x,y,0)=E_y(x,y,W_z)=0$ imposed by the field cage. However,  the term $\partial E_z/\partial z$ enters as well in 
 eq.~\ref{eq:Gauss-kappa}. Naturally, the boundary conditions are more effective in reducing the effects of space charge
 when both lateral dimensions are comparable to the gap length.
 Analytical approximations to $\kappa(x)$ are provided in  eqs.~\ref{eq:kappa-x-1}, \ref{eq:kappa-x-2} in appendix~\ref{app:analytical expressions}, together with approximations of the $y$ dependence of $E_y$. 

 \begin{figure}[bt]
\begin{center}
\includegraphics[width=0.9\textwidth]{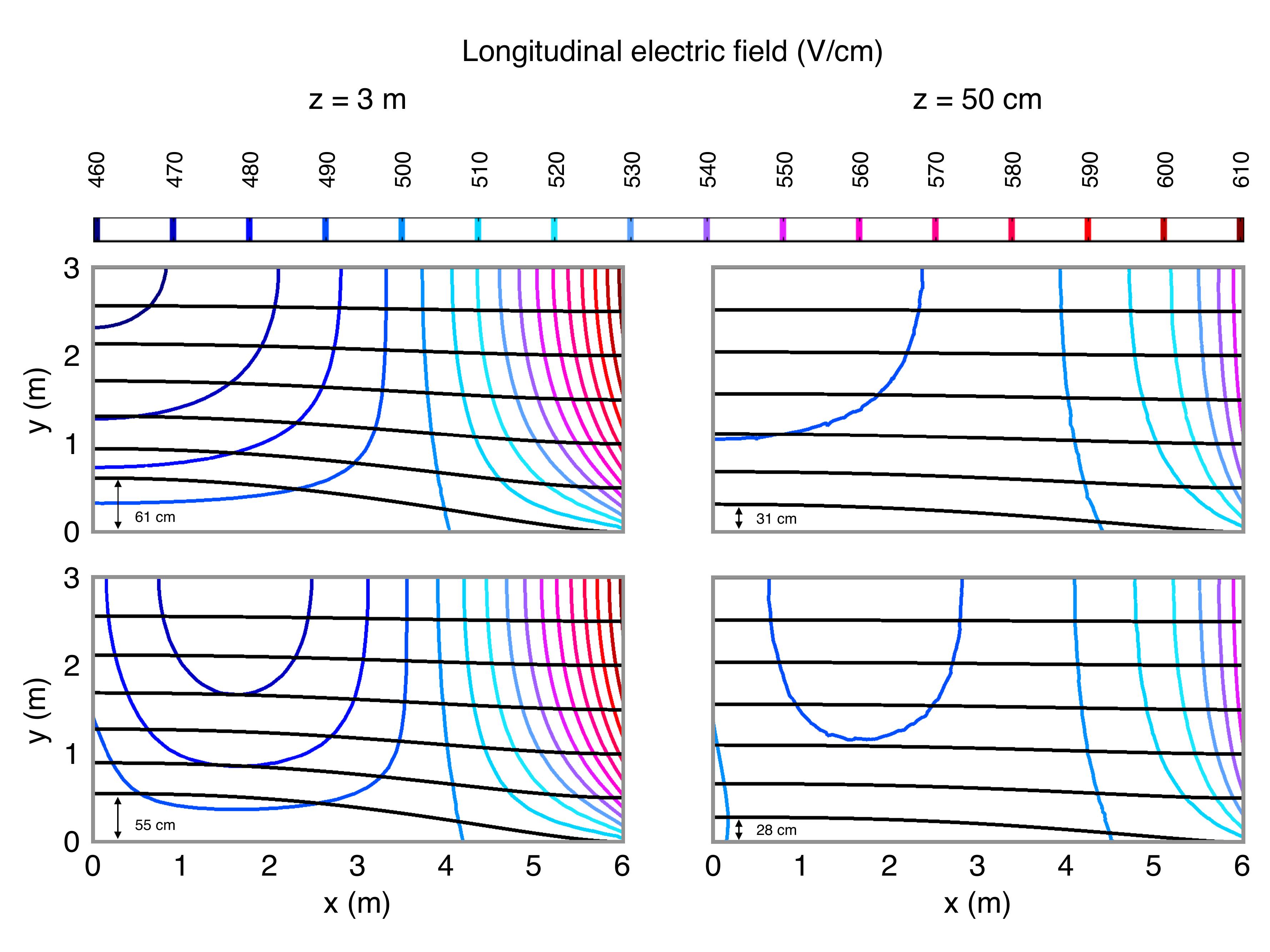}     
\caption {
Contours of equal value of $E_x$ and drift paths 
projected on the $(x,y)$ plane, for a cubic detector with 6~m sides, with $E_\circ = 500$~V/cm and $\alpha=1.15$.  
The plots on the top are obtained under the assumption of infinite electron lifetime, those on the bottom for 
$\tau_\text e = 10$~ms, to be compared to a drift time across the full gap of about 3.9~ms. 
The plot on the left are for the symmetry plane $z=3$~m, those on the right for $z=0.5$~m. }
\label{fig:3DEx-Paths-NC}
\end{center}
\end{figure} 

The numerical computation of the field strength along the drift direction is illustrated in figure~\ref{fig:3DEx-Paths-NC},
for a drift volume with $L=W_y=W_z=6$~m.
The contours of equal values of $E_x$ are shown on the symmetry plane $z = 3$~m, and on the plane $z=0.5$~m.
The plot for $z = 3$~m can be compared to figure~\ref{fig:2DExEy-tau}-left.  The proximity with the 
field cage in both $y$ and $z$ reduces the range of $E_x$ on the detector symmetry axis to about 460--610~V/cm, 
a factor 0.6 smaller than in the case of large detector widths.  
At a $z=0.5$~m the range of $E_x$ is reduced by an additional factor 0.5. 

The narrow aspect ratio has a smaller impact on on the lateral distortion, since  $\delta y$ is 
determined dominantly by the closest field cage. 
The comparison of figures~\ref{fig:3DEx-Paths-NC}-left and \ref{fig:2DExEy-tau} shows 
a 16\% reduction in $\delta y(x,0,z)$ at $z=3$~m.

 \begin{table}[tb]
\begin{center}
\caption {Comparison of different geometries of the drift volume: (a) $L \ll W_y\, , W_z$ and $\tau_e \gg 10$~ms; 
(b) $L \ll W_y\, , W_z$, $\tau_e=10$~ms;
(c) $L = W_y = W_z$, $\tau_e=10$~ms;
(d) $L = W_y = W_z$, $\tau_e=10$~ms and positive ions feedback parameter $\beta =1$. 
The results in columns (e), (f) refer to $L = W_y = W_z$ with the  correction to the field cage discussed in section~\ref{sec:mitig-field-cage}, with $\tau_e=10$~ms, and with $\beta = 1$ for column (f). 
The computation is made for $L=6$~m, $\alpha=1.15$.} 
\label{tab:aspect-ratio}
\begin{tabular}{| l | c c c c c c |}
\hline 
                                                                                    &   (a)  	&  (b)  	&   (c)      &     (d) 	& (e)		&   (f) 	\\
\hline                                     
$(E_x/E_\circ-1)_\text{min} \times\alpha^{-2} $           &  -17\%	&  -13\%	& -5\%   & -36\% 	 &	-9\%	&  -49\%	\\
$(E_x/E_\circ-1)_\text{max}\times\alpha^{-2}$          &  29\%   &  28\% 	& 17\%	& 29\% &    22\%   &  41\%	\\
$\left |E_y/E_\circ\right|_\text{max}\times \alpha^{-2}$&  18\% &  17\%	&	14\%	&  29\%	&	10\%	         &	22\%	 \\
$\delta y_\text{max}\times \alpha^{-2}L^{-1}$	         &  10\%&    9\%	&  7\%	& 16\% &    4.0\% 		&     11\% \\
\hline 
\end{tabular}
\label{Table:comparisons}
\end{center}
\end{table}

Table~\ref{tab:aspect-ratio} shows the comparison of different geometries, 
providing the extremes values for $E_x$ 
(on the detector axis, or far from the side walls), 
the maximum transverse component of the electric field $|E_y(x,0,z)|_\text{max}$, 
and the maximum transverse shift of an electron path $\delta_y(L,0,z)_\text{max}$. 
The numerical values are provided showing explicitly the dependence on the  
scaling variables $\alpha$, $L$, and can be applied to different configurations,  with relative accuracy of
some per cent for $\alpha < 1.5$.
The table also shows the result for a detector of equal sides $(L=W_y=W_z)$, including the case of positive ions feedback parameter $\beta =1$, and the effect of the mitigation procedure discussed in Sec~\ref{sec:mitig-field-cage}.

%


\section{Mitigation technique n.\ 2\,: correction to field cage}   \label{sec:mitig-field-cage}
The transverse components $E_y$, $E_z$ can be cancelled by means of setting the voltage gradient on the field cage $V_\text{fc}(x)$
so that it reproduces the voltage profile $V(x)$ corresponding to  the one-dimensional description of the effects of space charge. 
In that condition, the field cage is just as effective in removing boundary effects 
as the usual configuration of uniform gradient does for the case of   
$\alpha \ll 1$.  

However the ability to properly establish $V_\text{fc}(x)$ is limited by different factors. Firstly, one has to rely on an \emph{a priori} rather accurate knowledge of the ratio of charge density injection and ion mobility $K/\mu^+$, which includes the effect of ionisation yield and
determines the value of $\alpha$. A reasonable estimation of the electron lifetime $\tau_\text{e}$ is also needed, if $\lambda_e$ is not much larger than $L$. 
Secondly, as discussed below in section~\ref{sec:convection}, convective motion related 
to thermal gradients and to fluid recirculation may affect significantly the distribution of space charge, 
introducing additional dependences on $x, y$ and $z$,
and preventing an accurate cancellation of the transverse components of the electric field.

Because of these inherent difficulties, an approximate correction to the voltage profile of the field cage is discussed here:  besides the voltage imposed at the anode 
($V_\text{fc}(0)=0$) and cathode ($V_\text{fc}(L)=-V_\circ$), a third connection is provided to a resistive field-cage at a coordinate $x_\text{fc} \simeq L/2$, where the voltage is set a value  $V_\text{fc}$ with absolute value lower than 
$V_\circ \times (x_\text{fc}/L)$ and closer to the  voltage $V(x_\text{fc})$ observed at the same distance from the electrodes and far from the 
field cage.  The voltage profile on the field cage remains linear, but with a change of slope at $x=x_\text{fc}$.
Differently from the usual configuration, where the drifting electrons are attracted inward along the entire drift gap, and in particular for $x\simeq x_\text{fc}$, now the transverse component of the electric field is cancelled in the region where it was largest, and the remaining component  at $x \simeq x_\text{fc}/2$ and $x \simeq (x_\text{fc}+L)/2$ have smaller amplitude. 

The value of $V_\text{fc}$ can be considered as adjustable to the actual conditions of $\alpha$, $V_\circ$ and, to some extent, to the effects of convective motion in which the TPC is operated, in particular if independent values of 
$V_\text{fc}$ can be chosen for four sides of the field cage. 

\begin{figure}[t!]
\begin{center}
 \includegraphics[width=0.9\textwidth]{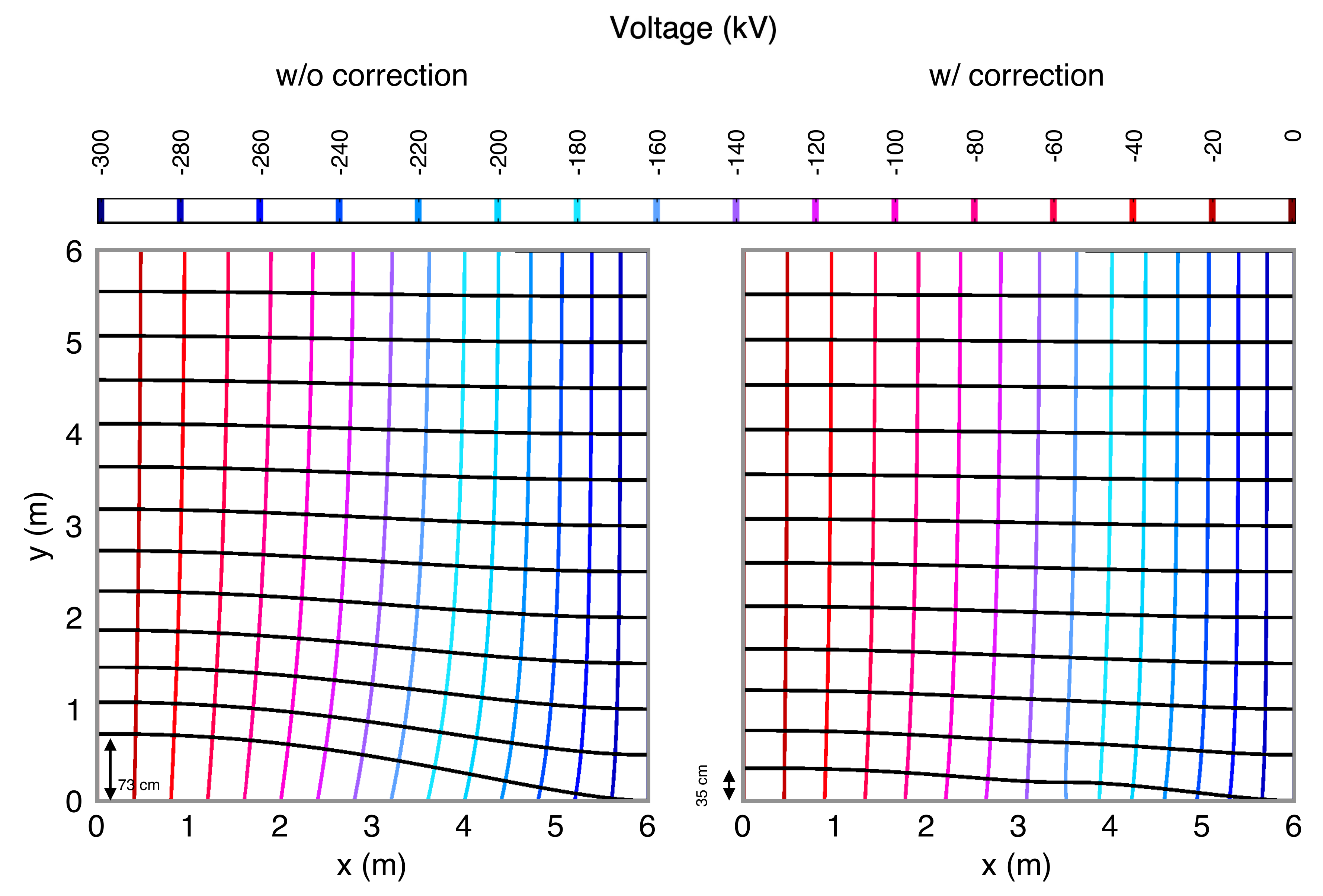}\\  
\includegraphics[width=0.9\textwidth]{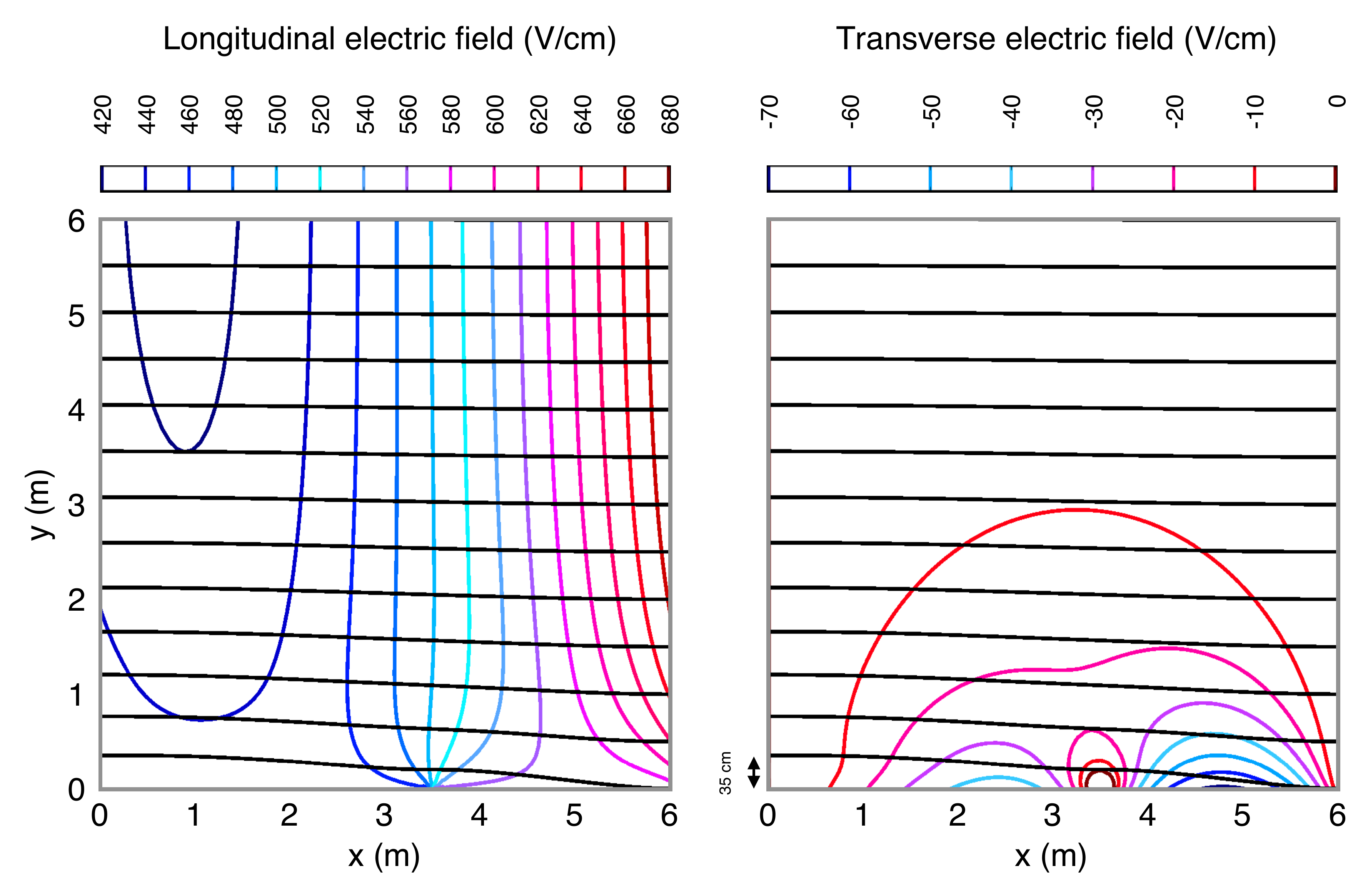}
\caption { 
Contours of equal voltage $V(x,y)$ and drift paths, with a usual voltage gradient at the field cage (\emph {top left}),
and with a third voltage connection at $x=3.5$~m, as discussed in the text (\emph {top right}). 
The plots on the bottom show the contours of equal $E_x(x,y)$ and equal $E_y(x,y)$ respectively, and drift paths, with the correction applied to the field cage.
The applied voltage is $V_\circ = 300$~kV, $\alpha=1.15$, and the width along $y$ is much longer than the gap length $L=6$~m.
}
\label{fig:2DEx-Paths-C}
\end{center}
\end{figure} 
Figure~\ref{fig:2DEx-Paths-C} shows the result of a numerical computation for a detector with $L=6$~m, $E_\circ=500$~V/cm, and large width, like in figure~\ref{fig:2DExEy-tau}. 
The $(x,y)$ projection of the equipotential contours and of the drift paths are shown for the usual field cage, and for  
a correction applied at $x=3.6$~m, where the voltage is set to $-159$~kV rather $-175$~kV. In this way, the transverse component of the electric field is locally cancelled.  The maximum transverse distortion of a drift path across the full gap is reduced by a factor 2, which applies also at larger $y$ values.  The corresponding contours of equal $E_x$ and $E_y$ are shown in the bottom plots, which may be compared to those of figure~\ref{fig:2DExEy-tau}.

\begin{figure}[t!]
\begin{center}
\includegraphics[width=0.9\textwidth]{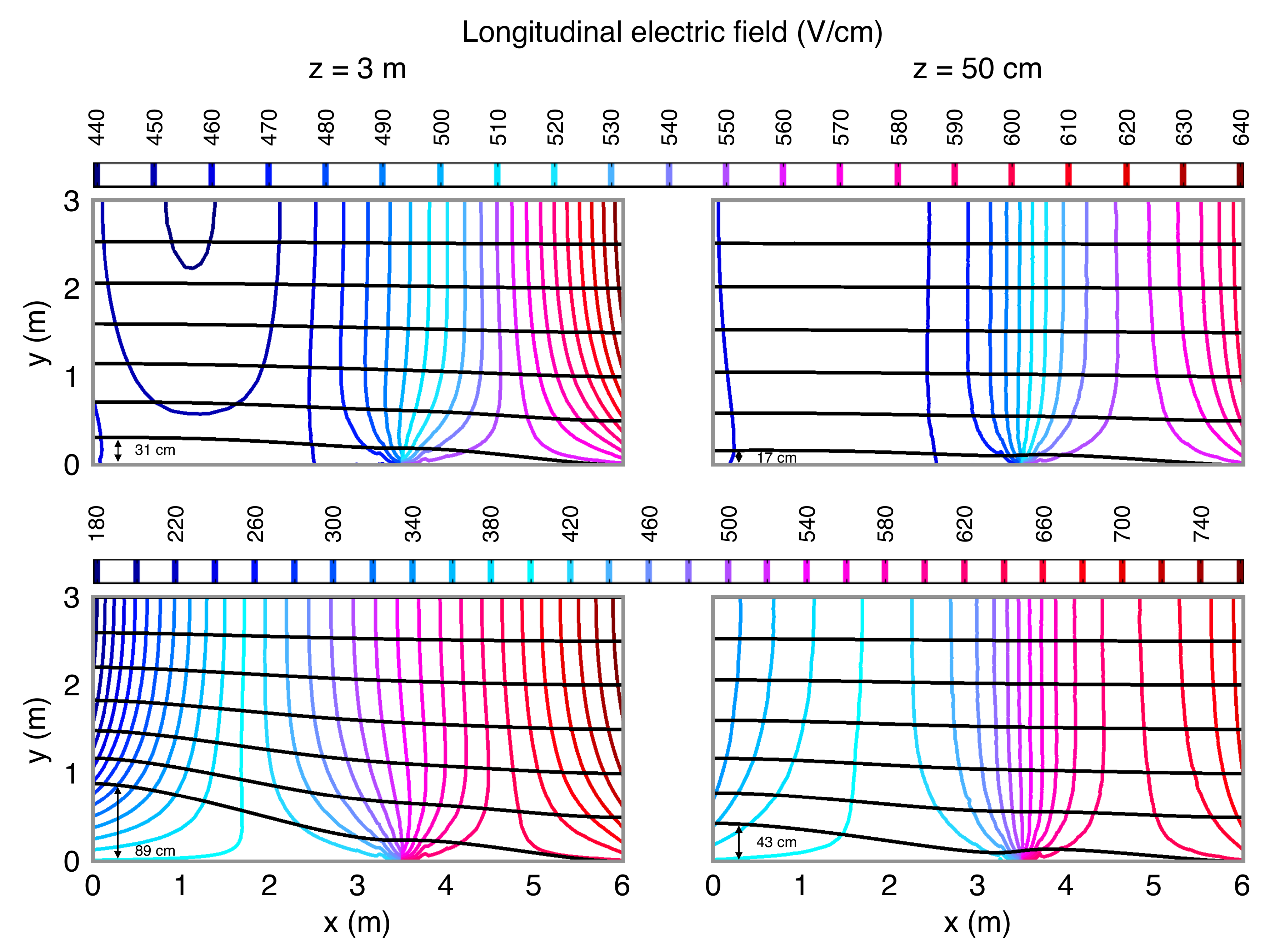}     
\caption {
Contours of equal value of $E_x$ and drift paths 
projected on the $(x,y)$ plane, for a cubic detector with 6 m sides, with $E_\circ = 500$~V/cm and $\alpha=1.15$,
with a correction applied to the field cage at $x=3.5$~m.
The plots on the bottom are computed with a positive ions feedback parameter $\beta=1$.
The plots on the left are computed for $z=3$~m, those on the right for $z=0.5$~m. }
\label{fig:3DEx-Paths-C}
\end{center}
\end{figure} 
\begin{figure}[h!]
\begin{center}
\includegraphics[width=0.9\textwidth]{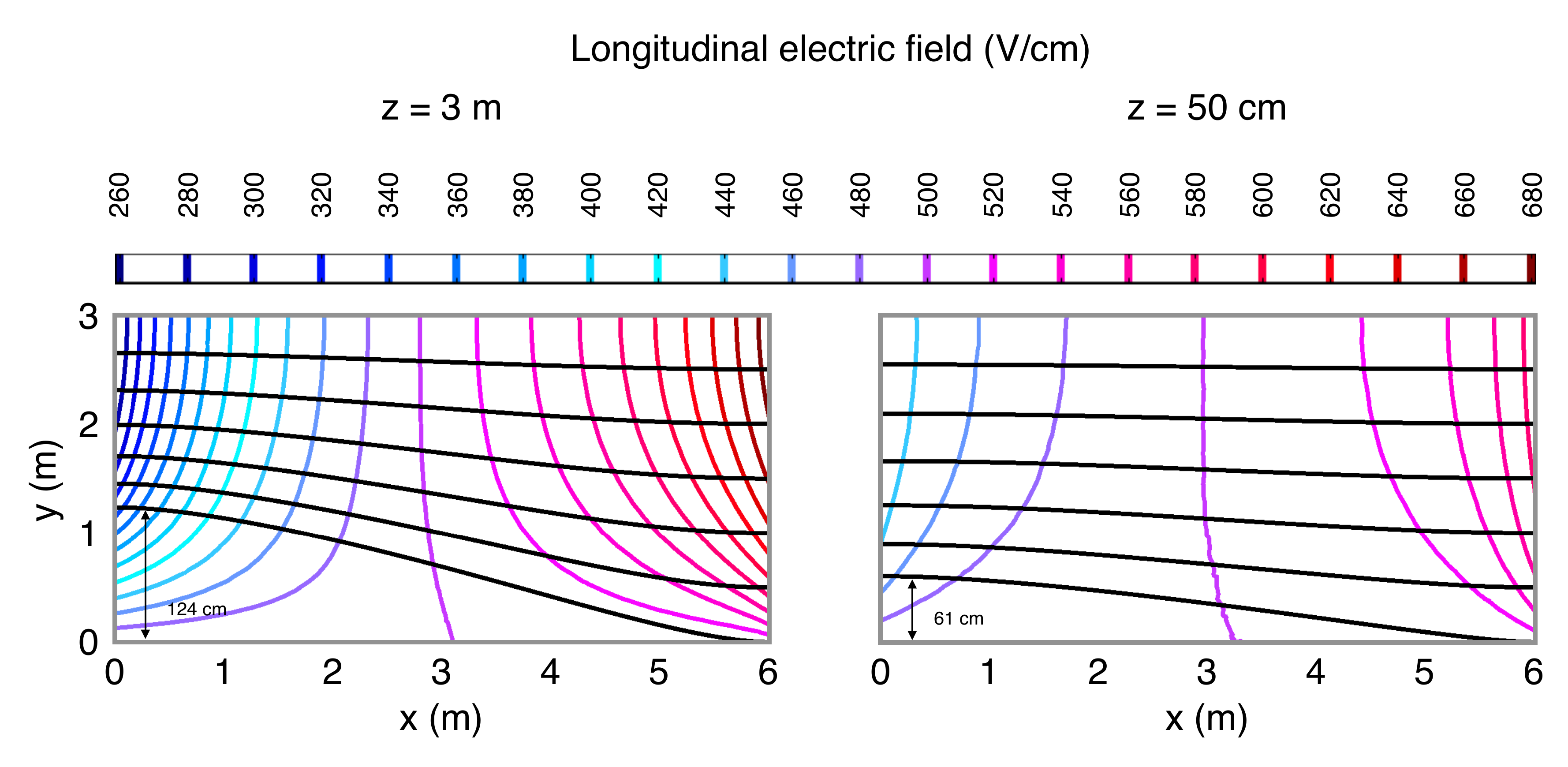}     
\caption {Contours of equal value of $E_x$ for a cubic detector, as in figure~\ref{fig:3DEx-Paths-NC}, with $\tau_\text e = 10$~ms 
and with positive ions feedback parameter $\beta=1$.}
\label{fig:3DEx-Paths-NC-beta}
\end{center}
\end{figure} 

For a narrow aspect ratio $L \lesssim W_y$ or $W_z$, as discussed in section~\ref{sec:aspect-ratio}  
the voltage profile established by the field cage affects the electric field across the entire detector volume. 
The correction to the field cage 
has the desirable effect of reducing the transverse components of the electric field,
achieving a more uniform detector response in the plane $y,\,z$. 
On the other hand, it increases the range of the $x$-dependent effects, with a larger reduction in $E_x$ in the region of the anode and a larger increase at the cathode. 
The effect is shown in figure~\ref{fig:3DEx-Paths-C}-top, which may be compared to figure~\ref{fig:3DEx-Paths-NC}.
A correction at $x=3.5$, with a reduction of the voltage from $-175$~kV to $-161$~kV, reduces the maximum transverse distortion by a factor 2,  but the range in the longitudinal component at $z = 3$~m is increased by a factor 1.5. 

A dual-phase detectors with feedback of positive ions requires larger corrections in order to cancel locally the transverse component of the electric field. Figure~\ref{fig:3DEx-Paths-C}-bottom shows the case with $\beta=1$, with the voltage at
$x=3.5$ raised to $-141$~kV. The comparison with figure~\ref{fig:3DEx-Paths-NC-beta}  shows a reduction in the maximum transverse distortion by a factor 0.71, but an increase in the range of variation of the longitudinal component of the electric field by a factor 1.4.
The comparisons among different configurations are summarised in Table~\ref{tab:aspect-ratio}.

In the examples shown in figures~\ref{fig:2DEx-Paths-C}--\ref{fig:3DEx-Paths-C},
the correction to the field cage is chosen aiming at a compensation of the transverse component of the electric field at
$x =x_\text{fc}$ and $y,\, z$ in proximity of the field cage. An over-compensation, with a larger correction to $V(x_\text{fc})$,
may be used to invert locally the sign of the transverse component, obtaining a further reduction of the value $\delta y_\text{max}$ for the full drift from anode to cathode. 
The limit to this option is posed the loss of drifting electrons from primary ionisation  for $x \gtrsim x_\text{fc}$ and 
sufficiently close to the field cage, which are driven outward and may be captured on the field cage without reaching the 
region $x \simeq  x_\text{fc}/2$, where they would be driven inward. 
If the active region at the anode starts at a distance $\Delta y$ from the field cage, a convenient design solution is to use a 
over-correction for which the maximum local outward displacement of drifting electrons matches the value of $\Delta y$.
 
Considerations on the optimal values for position $x_\text{fc}$ and the voltage $V(x_\text{fc})$ are presented in appendix~\ref{app:field-cage-corr}.




\section{Thermal convection effects}  \label{sec:convection}
The fluid dynamics of large liquid-argon TPCs are studied with numerical evaluations (see for instance \cite{convection}).
The pattern of the liquid flow is affected by evaporation at the surface, by thermal gradients induced by heat transfer at the cryostat walls, by heat dissipated in electronics contained in the liquid --- if present, and by liquid recirculation. The latter is performed for purification purposes, and contributes to fluid flow both directly and as an additional source of temperature non-uniformity.  
The value of the velocity field is typically predicted in the range of fractions of mm/s to several cm/s.   For comparison, the drift velocity of positive ions in typical liquid argon devices is in the range of 5--10~mm/s. Therefore the space-charge density distribution and the  pattern of longitudinal and transverse distortions may be altered in a significant way by fluid motion. 

Relatively large effects may be expected in proximity of elements that constrain the pattern of fluid motion, like side walls and possibly the electrode structures.  Asymmetries and anomalous behaviour of the transverse distortions $\delta y$, $\delta z$ near the field cages have been reported~\cite{Mooney:2015kke,ProtoDUNE-SP-performance}, suggesting the relevance of convective motion, 
while others~\cite{Antonello:2020qht} have observed that at some distance from the field cage, 
the observed longitudinal effects  $\delta x(x) $ are described in good approximation with the approach of
sections \ref{sec:1D-model} and \ref{sec:aspect-ratio}, without need to refer to liquid flow. 

It is not clear yet how accurate a fluid dynamics model cam be in predicting the distribution of space charge and the effects on drifting electrons, but studies are underway. Another desirable development would be a design of fluid recirculation, i.e.\ the geometry of inlets/outlets and the recirculation rate, that would take into account its influence on the distribution of ions, and minimise the related uncertainty in the prediction of space-charge effects. 

\section{Calibration strategies}
The discussion of calibration methods designed to correct for the effects of space charge lies outside the scope of this study. 
However, 
a description of usual calibration strategies is presented in this section.

Laser beams for calibration purposes have been used in large liquid argon TPCs \cite{MB-laser-calibration} or are being designed for future detectors 
\cite{DUNE-TDR-4}.  Two methods are usually considered. In the first, the laser is used to provide well defined sources of photoelectrons at predetermined locations. For example, photo targets placed on the cathode respond to the laser pulse providing a precise measurement of the total drift time and of the transverse distortion for drift paths across the full gap. 

In the second method, UV lasers are used to generated ionisation {\it tracks} in liquid argon TPCs via multi-photon ionisation \cite{laser-ionisation}. Movable mirrors can be used to steer laser beams across different regions of the drift volume,
suitable for comparison with uncalibrated tracks from reconstruction. 
The method is scarcely sensitive to the component of the distortion along the direction of the laser beam, 
but  crossing tracks from multiple laser units can be used 
to compare the expected coordinates of the crossing point with the corresponding apparent values, providing directly a local 
measurement of all components of the distortion.

A calibration based on crossing tracks has also been implemented using cosmic muons traversing a near-surface detector \cite{Abratenko:2020bbx}. 
The method relies on the observation that the end-points of tracks crossing the detector boundary are affected by space charge in a limited way. Indeed, after the timing of the data stream from a crossing muon is determined from the identification of the end-points on the anode or cathode, the coordinates of all end-points are constrained, since: 
(a) the space-charge effects vanish at the anode, and (b) on the side walls (field cage) the distortion occurs only along the direction normal to the wall, e.g.\ along $y$ in the notation of section \ref{sec:sidewalls}, since the components along the drift direction $x$ and and along $z$ are strongly suppressed in proximity of 
the field cage, where $E_z = 0$ and $E_x=E_\circ$, apart from local effects due to the granularity of the field shaping electrodes.
As discussed in sec.~\ref{sec:sidewalls}, the distortion $\delta y$ is directed towards the center of the detector and is equal to the apparent $y$ coordinate of the track end-point, and therefore can be directly determined from uncalibrated data, together with its dependence on $x$ and $z$.
Once all true end-points are determined, pairs of (nearly) crossing muons can be used to compare the expected coordinate of the  (near) intercept point with the corresponding one friom uncalibrated tracks.

\section{Conclusions}
The subject of space charge in large-size liquid argon TPC detectors has been reviewed, considering the effects 
on the longitudinal (drift time) and the transverse coordinates, and the implication on the measurement of the 
specific energy loss $dE/dX$. 
The subject is relevant for the important role that this detector technology is taking in present and future 
neutrino experiments.

Analytical description and numerical examples have been presented, displaying the dependence of the effects on 
detector size and operating conditions, and determining the dimensionless parameters that drive the behaviour of the detector response.
The potential enhancement of space-charge effects in dual-phase detectors with feedback of positive ions has been illustrated.
Boundary effects, and the case of detectors with comparable longitudinal and transverse size have been discussed. 

Two design solutions that mitigate the effects of space charge have been presented. In the simplest implementation, they
can reduce by at least a factor 2 the longitudinal and transverse distortions, respectively.  
The combination of the two solutions is a straightforward extension of the study presented here. 


%

\newpage

\appendix
\addcontentsline{toc}{section}{Appendices}
\section*{Appendices}

\section{Analytical expressions and approximations} \label{app:analytical expressions}
In the one-dimensional, basic model in which the electric field is described by eq.~\ref{eq:Ex}, 
the integral of $E_x(x)$ is given by:
\begin{equation}
V(x) \equiv -\int_0^x E_x(x^\prime)\,dx^\prime = 
       -\frac{x}{2}\, E_x(x) - \frac{L \, E_\text{a}{}^2}{2 \, \alpha \, E_\circ} \ln \left(\frac{E_x}{E_\text{a}} +\frac {\alpha \, x \, E_\circ}{L\, E_\text{a}}\right) \; .  \label{eq:V(x)-analytical}
 \end{equation}
The value of $E_\text{a}$ is determined by $V(L)=E_\circ L$. Figure~\ref{fig:Ea-Ec}  
(and figure~\ref{fig:backflow1Dcritical}, for the case without ion feedback) 
shows the result of a numerical computation of $E_\text{a}/E_\circ$ as a function of $\alpha$.
The following relations hold to first order in $\alpha^2$:
\begin{equation} \label{eq:Ea-first-order}
\begin{tabular}{r l l}
$E_\text{a} /E_\circ $ \hspace{-0.4cm}  &  $\simeq 1-\alpha^2/6 $  & better than 0.01 (0.05)  for $\alpha \! <  \! 1.2$ (1.6), \\
$E_\text{c} / E_\circ$ \hspace{-0.4cm}  & $ \simeq 1+\alpha^2/3 $ & better than 0.01 (0.05) for $\alpha  \! <  \! 0.75$ (1.12),   \\
$E_x(x) / E_\circ$ \hspace{-0.4cm}  &   $\simeq 1+(\alpha^2/2)(x^2/L^2-1/3) $ & better than 0.04 (0.14)  for $\alpha  \! <  \! 1$ (1.5) at any $x$,\\
$-V(x) /(E_\circ x)$ \hspace{-0.4cm}  &  $\simeq 1- (\alpha^2/6)(1-x^2/L^2) $  & better than  0.004 (0.03) for $\alpha \! <  \! 1$ (1.5) at any $x$.
\end{tabular}
\end{equation}

All these approximations fail as $\alpha$ approaches 2.
The maximum deviation between $V(x)$ and $-E_\circ \, x$ is
\begin{equation}
\delta V_\text{max}  =  [V(x) - E_\circ\, x]_\text{max}  \simeq    \frac{\alpha^2  E_\circ\, L}{16} \; ,   \label{eq:dVmax} \nonumber
\end{equation}
which occurs at $x_\text{max}$ between $L/\sqrt{3}$ and $L/2$ for increasing values of $\alpha$. 

Approximations at higher order in $\alpha^2$ have been obtained numerically:
\begin{equation}   \label{eq:Ea-higher-order}
\begin{tabular}{r l l}
$E_\text{a} / E_\circ $ \hspace{-0.4cm}  &  $ \simeq (1-\alpha^2/6-\alpha^4/180)$  
                                                                  & better than 0.01 (0.05) for $\alpha<1.57$ (1.82),  \\
                                     \hspace{-0.4cm}  &  $\simeq (1-\alpha^2/6-\alpha^4/180-\alpha^{10}/8500) $  
                                                                  & better than 0.01 (0.05) for $\alpha<1.89$ (1.97).  \\
$E_\text{c} / E_\circ  $ \hspace{-0.4cm}  &  $\simeq (1+\alpha^2/3-\alpha^4/30)$  
                                                                   & better than 0.01 (0.05) for $\alpha<1.30$ (1.60), \\
                                    \hspace{-0.4cm}  &  $\simeq  (1+\alpha^2/3-\alpha^4/29+x^6/360)  $     
                                                                   & better than 0.01 (0.05) for $\alpha<1.59$ (2.00).    
\end{tabular}
\end{equation}
 
At the field cage ($y=0$) of a detector of large widths $W_y,\: W_z \gg L$, and for $z$ far from the field cage, the transverse component of the electric field is approximated by the expression
\begin{equation} \label{eq:Eymax}
|E_y(x,0)| \simeq [1.17+25.7\,x/L+55.1\,(x/L)^2-80.7\,(x/L)^3]\times(\alpha^2 E_\circ/100) \; ,
\end{equation}
with an accuracy of some per cent of the maximum value of $|E_y(x,0)|$.
 
 The $x$ dependence of the factor $\kappa(x)$ describing  $\partial E_y/\partial y \,(x)$ at $y=W_y/2$, $z=W_z/2$ is given with an accuracy of a few per cent by the expressions:
\begin{eqnarray} 
\kappa (x) = \: 0.53 \; \alpha^2 \, \frac{E_\circ }{L} \, \frac{x}{L} \left[ 1- \left( \frac{x}{L} \right) ^2 \right] \quad & \text{for } \, L=W_y \ll W_z \; , 
\label{eq:kappa-x-1}  \\
\kappa (x) = \: 0.41 \; \alpha^2 \, \frac{E_\circ }{L} \; \frac{x}{L} \left[ 1- \left( \frac{x}{L} \right) ^2 \right]  \quad & \text{for } \, L=W_y =W_z \; .\label{eq:kappa-x-2}
\end{eqnarray}

For $W_y \gg L$, $W_z \gg L$, $L \ll z \ll (W_z-L)$, $x \simeq L/2$ and $y \ll W_y/2$,  the $y$ dependence of $E_y(x,y)$ is approximated with an accuracy of a few per cent by
\begin{equation}  \label{eq:EyL2y}
E_y(x,y) \simeq E_y(x,0) \times \exp(-2.56\,y/L -0.32\,y^2/L^2) \equiv
                          E_y(x,0)\times f_{x=L/2}(y)   \:.   \nonumber
\end{equation}
As discussed in section~\ref{sec:aspect-ratio}, as $y$ increases towards  $W_y/2$, the effect of the field cage at 
$y=W_y$ becomes more relevant, determining $E_y(x,W_y/2)=0$.  The combined effect of the two boundaries 
is well described by 
\begin{equation}
E_y(x,y) \, \simeq \, E_y(x, 0) \times [ f_x(y) - f_x(W_y-y) ]    \qquad (0 \le y \le  W_y)\, .\nonumber
\end{equation}
This equation is valid at the level of few per cent relative accuracy for $L=6$~m, $0 \le x \le L$ 
and $W_y=12$~m, and better than 10\% for $W_y=6$~m.

\section{Analytical approximation with finite electron lifetime in one dimension}   \label{app:1D-lifetime}
In one dimension (i.e., far from the side walls)
the set of equations describing the electric field and the charge density distributions for 
electron, positive and negative ions, including a finite electron lifetime, is:
\begin{eqnarray}
\frac{dE}{dx}&=&\frac{\rho^+ + \rho^-}{\varepsilon}   \nonumber \\
\frac{d(\rho^\text{e}v^\text{e})}{dx}&=&-K-\frac{\rho^\text{e}}{\tau_\text{e}} \nonumber \\
\frac{d(\rho^+v^+)}{dx}&=&K  \nonumber  \\
\frac{d(\rho^-v^-)}{dx}&=&\frac{\rho^\text{e}}{\tau_\text{e}} \nonumber 
\end{eqnarray}
where $\rho^-$, $\rho^\text{e}$, $v^-$ and $v^\text{e}$ are negative, and $\rho^\text e$ is negligible when compared to
$\rho^+$ and $\rho^-$.
 The dependence of $v^\text{e}$ on $E$ prevents a direct integration. An approximate solution can be found under the assumptions:
\begin{equation}
\frac{|\rho^\text{e}|}{\tau_\text{e}}\ll K \; , \qquad |v^\text{e}(E)| \simeq v^\text{e}(E_\circ)\equiv v^\text{e}_\circ  \nonumber \; ,
\end{equation}
which lead to $\rho^\text{e}\simeq -K(L-x)/v^\text{e}_\circ $, 
a reasonable approximation of the numerical solution shown in figure~\ref{fig:1D-abso}. 
The equations for $\rho^+$, $\rho^-$ and $E$ can then be integrated directly as
\begin{eqnarray}
\rho^+(x)&\simeq&\frac{K\,x}{\mu^+ E(x)} \nonumber  \\
\rho^-(x)&\simeq&-\frac{K(L-x)^2}{2\,\lambda_\circ^\text{e}\mu^-E(x)} \nonumber \\
E(x)&\simeq&E_\circ\sqrt{ \left(\frac{E_\text{a}}{E_\circ}\right)^2 
    +\alpha_+^2\left(\frac{x}{L}\right)^2 
    -\alpha_-^2 \left(\frac{L}{\lambda^\text{e}_\circ}\right) \left(\frac{x}{L}-\frac{x^2}{L^2}+\frac{x^3}{3\,L^3} \right)
    } \nonumber
\end{eqnarray}
where $\lambda^\text{e}_\circ=v^\text{e}_\circ \tau_\text{e}$.
In the comparison with the numerical evaluation illustrated in figure~\ref{fig:1D-abso}, with $\alpha_+=\alpha_- =1.15$ and 
$\lambda^\text{e}_\circ /L=2.58$, 
the analytical approximation is found to be accurate to 1\% in the values of 
$E(x)$ and $\rho^+(x)$, 
and in the range (10--15)\% in $\rho^-(x)$. 
For $\lambda^\text{e}_\circ /L=1.29$, the accuracy is better than 2\% and (10--30)\% respectively.

\section{Electric field with separation grid and optimal configuration}
\label{app:1D-grid}
As discussed in section~\ref{sec:grid}, with a separation grid placed at $x_\text{g}$ with voltage $V_\text{g}=-\delta_\text{g} V_\circ = -(x_\text{g}/L)V_\circ$, the electric field for $x \le x_\text{g}$ is still given by equation~\ref{eq:Ex}:
\begin{equation}
E_x(x) =  E_\circ \sqrt{(E_\text{a}/E_\circ)^2 + \alpha^2 (x/L)^2 } \qquad (x < x_\text{g}) \;,  \nonumber 
\end{equation}
with the only difference that the value of the electric field at the anode $E_\text{a}$ is derived from the dependence of $E_\text{a}$ vs.\ $\alpha$, 
shown in figure~\ref{fig:Ea-Ec}  and in eqs.~\ref{eq:Ea-first-order}, \ref{eq:Ea-higher-order},
after replacing $\alpha$ with $\delta_\text{g}\alpha$. 
Since at lowest order the effects of space charge are proportional to $\alpha^2$, 
in the region between the anode and the grid the distortion to the electric field is reduced by a factor $\delta_\text{g}{}^2$. The
value of $\delta_\text g$ cannot be too small, though, in order to preserve favourable effects of the grid on the side $x > x_\text g$.

In the region $x \ge x_\text{g}$, the boundary condition on the flux of positive ions is  
\begin{equation}
(\rho^+v_x^+)_{x=x_\text{g+}} = (\rho^+v_x^+)_{x=x_\text{g-}}\times (E_\text{g+}/E_\text{g-}) = K\,x_\text{g}\, (E_\text{g+}/E_\text{g-})\:, \nonumber 
\end{equation} 
where $K\,x_\text{g}$ is the ions flux reaching the grid from the $x < x_\text g$ side, which crosses the grid at a fraction equal to the ratio of electric field on the two sides. 
The continuity equation provides
\begin{equation}
\rho^+(x) = \frac{K\,(x-x_\text{g}) + K\,x_\text{g}  (E_\text{g+}/E_\text{g-})}{\mu^+\,E_x(x)} \qquad (x > x_\text{g}) \: ,
\nonumber 
\end{equation}
and the corresponding differential equation for $E_x(x)$ is directly integrated as given in Eq.~\ref{eq:grid-E-cathode}.

The optimal grid position $x_\text g$ may be chosen so that in the corresponding case without grid, 
the field variation between $x=0$ (anode ) and $x=x_\text{g}$ is equal to half of the total variation across the gap, so that the range of 
variation in $E_x$ in the two regions 
is approximately equal. At first order in $\alpha^2$, the position determined in this way is 
$x_\text g \simeq  (1/\sqrt{2})(1-\alpha/16)$.  Next, the value of $\delta_g$ is chosen so that the range of $E_x$ is equal in the two region, namely: $E_\text{a}=E_\text{g+}$, $E_\text{g-} = E_\text{c}$. 

As an example, for $\alpha=1.6$, the parameters of the optimal configuration are shown in first line of Table~\ref{tab:grid}, together
with the corresponding values of $E_x$ at the electrodes. The full range 
$(E_{x|\text{max}}-E_{x|\text{min}})/E_\circ$ is 0.503, to be compared to 1.161 without the grid. 
The closeness between the values $x_\text g /L=0.642$ and $\delta_\text g=0.626$ suggests to consider the configuration with equal values,  which is shown in the third line of the table, and for which the range in $E_x/E_\circ$ is increased by 0.052 only, and the lowest value of $E_x/E_\circ$ is reduced by 0.034,  which is also a small value, when compared to the difference of 0.270 obtained with the configuration without grid. Therefore the condition $x_\text g /L = \delta_\text g$ does not affect significantly the optimisation of the separation grid. 

Additional examples that approach optimisation within the constraint $x_\text g /L = \delta_\text g$ are shown in other lines of Table~\ref{tab:grid}, 
together with a set of values obtained for 
$x_\text g/L = \delta_\text g = 0.7$, which are used in figure~\ref{fig:grid}.  
For $\alpha=2$, the ranges of $E_{x}/E_\circ$ is 14\% wider with  $x_\text g/L$ equal to 0.7 rather than 0.6, but the 
the values $E_{x|\text{min}}/E_\circ$ are similar, 0.63 and 0.65 respectively, so that in the end the effectiveness of the separation grid does not depend much on the choice of the value $x_\text g/L$ within the range 0.6 to 0.7. 
\begin{table}[hbt]
\begin{center}
\caption {Examples of configurations of separation grid and resulting values of the electric field at the electrodes.
The last three lines refer to the case shown in figure~\ref{fig:grid}.}
\label{tab:grid}
$\begin{array}{ | c | c c c c c c |}
\hline 
\alpha &  x_\text g /L & \delta_\text g &  E_\text{a} /E_\circ &  E_\text{g-} /E_\circ& E_\text{g+} /E_\circ& E_\text{c}/E_\circ  \\                                                    
\hline                                     
 \hphantom{a} 1.6 \hphantom{a} & 0.642 & 0.626 & 0.792 & 1.295 & 0.792 & 1.295 \\
 \hline
 0.8 & 0.68   &  0.68  & 0.953   &  1.097  & 0.925   & 1.075 \\
 1.6 & 0.64   &  0.64  & 0.822    & 1.313  & 0.758  & 1.260 \\
 2.0 & 0.60   &  0.60  & 0.748   &  1.414 & 0.630 & 1.376 \\
 \hline
 0.8 & 0.7 & 0.7 & 0.950 & 1.103 & 0.925 & 1.067 \\
 1.6 & 0.7 & 0.7 & 0.783 & 1.366 & 0.790 & 1.215 \\
 2.0 & 0.7 & 0.7 & 0.650 & 1.544 & 0.716 & 1.285 \\
\hline 
\end{array}$
\end{center}
\end{table}

\section{Parameters for field cage correction}
\label{app:field-cage-corr}
For a detector with wide aspect ratio $L \ll W_y\, , W_z$, and considering only positive ions, the voltage difference $\delta V_\text{fc}(x)$ between a point of drift coordinate $x$ 
and transverse coordinates $y$, $z$ far from the field cage and the corresponding voltage at the position $x$ of a linear field-cage is given by the voltage distortion $\delta V(x)$ in the one-dimensional description of the effects of space charge. 
Near the field cage, the voltage difference generates transverse components of the electric field according to Eq.~\ref{eq:Et-int}, and 
the most effective position $x_\text{fc}$ for a correction to the voltage profile of the field cage is at the maximum of $\delta V_\text{fc}(x)$.

For $\alpha < 1.5$, 
$\delta V_\text{fc}(x) = V(x)+E_\circ\,x$ is approximated using eq.~\ref{eq:Ea-first-order}, and the preferred coordinate is 
$x_\text{fc} \simeq L/\sqrt{3}$. A correction voltage 
$V_\text{fc}= - (E_\circ\,L /\sqrt{3})(1-\alpha^2/9)$ is applied to the field cage at that point in order to cancel $\delta V_\text{fc}$ locally 
and reduce it in the full range $0 < x < L$. 
Compared to a field cage without correction, the current flowing from the anode to $x_\text{fc}$ is 
reduced by a fraction equal to $\alpha^2/9$, the current flowing from $x_\text{fc}$ to the cathode is increased by a fraction equal to
$\alpha^2/[9(\sqrt{3}-1)]$, with the difference of about $0.26\,\alpha^2$ being provided at the intermediate connection. 

Once the correction is applied, $\delta V_\text{fc}(x)$ takes a different shape and two local maxima occur 
at $x=(1/3)L$ and $x\simeq0.80\,L$.  The values of the maxima are equal to 
$\alpha^2 E_\circ\, L /81$ and about $\alpha^2 E_\circ\, L /57$ respectively, much smaller than $\alpha^2 E_\circ\, L /16$ 
obtained without correction to the field cage.

For larger values of $\alpha$, without applying the correction, the maximum of $\delta V_\text{fc}(x)$ moves towards $x=L/2$. For instance, for $\alpha=1.5$, the preferred 
value is $x_\text{fc}\simeq 0.54\, L$, with the maximum of $\delta V_\text{fc}(x)$ still approximated by $\alpha^2 E_\circ\, L /16$ within about 1\%.

When negative ions from electron capture are present, smaller values of $\delta V_\text{fc}(x)$ are found for the same value of $\alpha$.
Computation performed with $\tau_\text{e}=10$~ms, $L=6$~m, $E_\circ=500$~V/cm and $\alpha = 1.15$, 
have shown that, without the correction, $\delta V_\text{fc}(x)$ is about 10\% smaller than in in the case with infinite electron lifetime, with the
coordinate at the maximum of $\delta V_\text{fc}$ still well approximated by $x \simeq L/\sqrt{3}$.

\end{document}